\documentclass{siamltex}

\usepackage{cite}
\usepackage{subeqn}
\usepackage{graphicx}
\usepackage{epsfig}
\usepackage{amssymb}

\newcommand{\lewis}{{\mbox{Le}}}

\def\CwedgeX{\mbox{$\bigwedge^{2}({\mathbb C}^{4})$}}
\def\R{ {\mathbb R} }

\def\C{ {\mathbb C} }
\def\Z{ {\mathbb Z} }

\def\lbk{\lbrack\!\lbrack}
\def\rbk{\rbrack\!\rbrack}
\def\re{{\rm e}}
\def\fr {\mbox{$\frac{1}{2}$}}

\begin{document}

\title{High Lewis number combustion wavefronts: a perturbative Melnikov analysis}

\author{
Sanjeeva Balasuriya\thanks{[Corresponding author] Department of Mathematics,
Connecticut College, 270 Mohegan Avenue, New London, CT 06320, USA
({\tt sanjeeva.balasuriya@conncoll.edu}).} \and
Georg Gottwald\thanks{School of Mathematics \& Statistics,
University of Sydney, NSW 2006, Australia
({\tt gottwald@maths.usyd.edu.au}).} \and
John Hornibrook\thanks{School of Mathematics \& Statistics,
University of Sydney, NSW 2006, Australia
({\tt johnh@maths.usyd.edu.au}).} \and
St\'ephane Lafortune\thanks{Department of Mathematics, College of Charleston,
66
George Street,
Charleston, SC 29424, USA
({\tt lafortunes@cofc.edu}).}
}

\maketitle

\begin{abstract}
The wavefronts associated with a
one-dimensional combustion model with Arrhenius kinetics and no heat
loss are analyzed within the high Lewis number perturbative limit.
This situation, in which fuel diffusivity is
small in comparison to that of heat, is appropriate for highly dense
fluids. A formula for the wavespeed is
established by a non-standard application of Melnikov's method
and slow manifold theory from dynamical systems, and compared to
numerical results.  A simple characterization of the wavespeed correction
is obtained: it is proportional to the ratio between the exothermicity
parameter and the Lewis number.  The
perturbation method developed herein is also applicable to more general
coupled reaction-diffusion equations with strongly differing
diffusivities.  The stability of the wavefronts is also tested
using a numerical Evans function method.
\end{abstract}



\pagestyle{myheadings}
\thispagestyle{plain}
\markboth{BALASURIYA, GOTTWALD, HORNIBROOK AND LAFORTUNE}{HIGH LEWIS NUMBER COMBUSTION WAVEFRONTS}

\section{Introduction}
\label{sec:intro}

In this article, we study the wavespeed of a combustion
wavefront along a one-dimensional medium.  This is a fundamental
idealized problem towards understanding how flame fronts propagate,
and therefore has received a considerable amount of attention.  There
are several (non-dimensional) parameters of
importance: the Lewis number $ \lewis $, the exothermicity parameter $
\beta $, and the heat loss parameter $ \ell $.  The first of these,
the Lewis number, measures the relative importance of fuel diffusivity
in comparison to that of heat.  The exothermicity $ \beta $ is the
ratio of the activation energy to the heat of reaction.  The structure
of the governing equations is such that an infinite Lewis number
is considerably easier to deal with than allowing for fuel
diffusivity.  Many studies of this ``solid'' regime appear in the
literature \cite{vv,WeberLe, ms,bill,bmtwo}, and also the ``gaseous''
regime $ \lewis \approx 1 $ has been frequently studied because of a
symmetry in the equations
\cite{WeberLe,EvansWeberGroup,bill,kapila,xin}.  Usually, the heat
loss is neglected in these ``adiabatic'' studies.  In several of these
articles \cite{WeberLe,vv,ms} the condition $ \beta
\gg 1 $ is essential to the wavespeed and stability analysis.  The
case $ \beta \ll 1 $ has also been studied \cite{bmexo}, in which a
perturbative method is used to model the temperature.  The bifurcation
structure with respect to the heat loss parameter $ \ell $ is
addressed in \cite{skms}, which obtains a stability diagram with
respect to $ \ell $ and the wavespeed.

We note that the limit of small fuel diffusivity (large, but not
infinite, Lewis number) has not received much attention, perhaps
because of the singularity of this limit in the governing equations.
Yet this limit may be argued to be particularly appropriate for very
high density fluids burning at high temperatures, such as would occur,
for example, in the burning of toxic wastes at supercritical
temperatures \cite{mj}.  Even for solids, some mass diffusivity is to
be expected at very high temperatures, particularly in the reaction
zone in which liquification may occur. In \cite{margolis91}, the mass
diffusivity is modeled by an Arrhenius temperature dependence, which
would result in a large effective Lewis number in certain situations
(such as when the [scaled] adiabatic flame temperature is small in
comparison to the activation energy for mass diffusion).  It is this
very large Lewis number limit which we study in this article, without
restricting $ \beta $. We do a detailed analysis of the wavespeed of
combustion waves which can be supported.  We also verify the linear stability
of such wavefronts using an Evans function technique.

The model we use is for a premixed fuel in one dimension, with no heat
loss and with an Arrhenius law for the reaction rate.  These
combustion dynamics can be represented in non-dimensional form by
\cite{WeberLe,EvansWeberGroup,skms,bmtwo,bmexo,vv,ms,kapila,xin}
\begin{equation}
\label{eq:governpde}
\renewcommand{\arraystretch}{1.5}
\left\{ \begin{array}{lll}
\displaystyle \frac{\partial u}{\partial t} & = & \displaystyle \frac{\partial^2 u}{\partial x^2}
+ y \, e^{-1/u}  \\
\displaystyle \frac{\partial y}{\partial t} & = & \displaystyle \frac{1}{\lewis}
\, \frac{\partial^2 y}{\partial x^2} - \beta \, y \, e^{-1/u}
\end{array} \right. \, .
\end{equation}
Here, $ u(x,t) $ is the temperature, and $ y(x,t) $ the fuel
concentration, at a point $ x $ at time $ t $.  The parameters $ \beta
$ and $ \lewis $ are as described earlier.  We are neglecting heat
loss (had we included it, an additional term $ - \ell \left( u - u_a
\right) $ for some ambient temperature $ u_a $
would be necessary on the right-hand side of the $ u $ equation in
(\ref{eq:governpde})).  This one-dimensional model is also applicable
to combustion in cylinders \cite{2Dand1D}, with $ u $ and $ y $ being
cross-sectionally averaged quantities in this case.  See also
\cite{gklmms,mw,bmtravel,bill} for closely related governing
equations. The non-dimensionalization leading to (\ref{eq:governpde})
ensures that the cold boundary problem is circumvented (see \cite{vv}
for a discussion).  Since the Lewis number will be assumed large, set
$ \epsilon = 1 / \lewis $ with $ 0 \le \epsilon \ll 1 $.
This small $ \epsilon $ limit clearly constitutes a singular
perturbation in (\ref{eq:governpde}).

This article analyzes (\ref{eq:governpde}) as follows.  In
Section~\ref{sec:wavespeed}, we determine the wavespeed as a function
of $ \beta $ and $ \epsilon $.  We initially consider the situation
where $ \lewis = \infty $ (Section~\ref{sec:infty}), since this
wavespeed is relevant to our subsequent perturbative analysis for $ 1
\ll \lewis < \infty $ (Sections~\ref{sec:large}, \ref{sec:reduce} and \ref{sec:mel}).
While the infinite Lewis number situation is well-studied, we are able
to empirically determine a simple exponential formula for the wavespeed as
a function of $ \beta $.  The case $ 1 \ll \lewis < \infty $ is initially
examined numerically in Section~\ref{sec:large}, in which we obtain a
method for computing the wavespeed.  In the subsequent sections, we establish
a theoretical estimate for the wavespeed with the help of two suitably
modified tools from dynamical systems theory: a slow manifold reduction,
and Melnikov's method.  In Section~\ref{sec:reduce}, we reduce the dimensionality
of the problem using a
slow manifold reduction argument.  This enables us
in Section~\ref{sec:mel} to utilize a non-standard adaptation of Melnikov's
method to find a theoretical estimate for the wavespeed.
(This new technique is adaptable to other situations in
which the wavespeed correction due to the presence of a small
parameter is needed.)  Our asymptotics enable the determination of a
remarkably simple formula for the
wavespeed, which is accurate for {\em all} $ \beta $ values (and not
restricted to the ``usual'' large $ \beta $ limit).  Essentially, we find
that the relative wavespeed correction in going from infinite to large
Lewis number is proportional to $ ( \beta / \lewis) $.

A brief stability analysis of the wavefronts is given in
Section~\ref{sec:stability}.  Having described the Evans function approach to stability in
Section~\ref{sec:stability}, we compute the Evans function for high Lewis
number combustion wavefronts using an exterior algebra \cite{Ev75,Jo84,Ya85,Al90}.  We
note that in \cite{EvansWeberGroup}, an exterior algebra method has been successfully
used to numerically investigate stability of wavefronts in combustion systems.
A detailed stability analysis in the $ \beta $ - $ \lewis^{-1} $ plane is given therein
for the system (\ref{eq:governpde}).
As with infinite Lewis number fronts (see \cite{WeberLe,Cardarelli,ms,bmtwo}), \cite{EvansWeberGroup} show that
stability occurs for small $ \beta $, but that as $ \beta $ is increased, a Hopf
bifurcation leads to an oscillatory instability.  The $ \beta $ and $ \lewis $
values we test give results consistent with the stability boundary determined
in \cite{EvansWeberGroup}.  Thus, stability properties remain essentially unaltered
despite the singularity in the limit $ \lewis \rightarrow \infty $.

\section{Wavespeed analysis}
\label{sec:wavespeed}

We seek wavefronts which travel in time, and hence set $ u(x,t) =
u(\xi) $ and $ y(x,t) = y(\xi) $, where $ \xi = x - c \, t $ and $ c $
is the traveling wave speed.  Under this ansatz, (\ref{eq:governpde})
reduces to
\begin{equation}
\label{eq:governode}
\left\{
\begin{array}{lll}
u^{\prime \prime} + c \, u^{\prime} + y \, e^{-1/u}  & = & 0 \\
\epsilon \, y^{\prime \prime} + c \, y^{\prime} - \beta \, y \, e^{-1/u} & = & 0
\end{array} \right. \, .
\end{equation}

\subsection{Wavefront for $ \lewis = \infty $}
\label{sec:infty}

Set $ \epsilon = 0 $ in (\ref{eq:governode}). Upon defining the new
variable $ v = u^\prime $, the dynamics can be represented by a
three-dimensional first-order system
\begin{equation}
\label{eq:ode0}
\left\{
\renewcommand{\arraystretch}{1.3}
\begin{array}{lll}
u^\prime & = & \displaystyle v \\
v^\prime & = & \displaystyle - c \, v - y \, e^{-1/u} \\
y^\prime & = & \displaystyle \frac{\beta}{c} \, y \, e^{-1/u}
\end{array}
\right. \, .
\end{equation}
The system (\ref{eq:ode0}) possesses a conserved quantity
\begin{equation}
H_c(u,v,y) = \beta \, v + \beta \, c \, u + c \, y \, ,
\label{Hc3}
\end{equation}
since it is verifiable that $ d H_c / d \xi = 0 $ along trajectories
of (\ref{eq:ode0}).  Thus, motion is confined to planes defined by $
H_c(u,v,y) = $ constant.  Now, for a wavefront, we require that $
(u,v,y) \rightarrow (0,0,1) $ as $ \xi \rightarrow \infty $; this
corresponds to the region in which fuel is not yet burnt (and remains
at its maximum non-dimensional concentration of one) and the temperature (and
its variation) is still zero.  This point lies on $ H_c(u,v,y) = c $,
giving a well-known conservation relation \cite{WeberLe}.  At the
other limit $ \xi \rightarrow - \infty $, the fuel is completely
burnt, and has reached a steady temperature, and so $ (u,v,y)
\rightarrow (u_B,0,0) $, where the temperature $ u_B $ is to be
determined.  Utilizing $ H_c(u_B,0,0) = c $, we find that $ u_B = 1 /
\beta $ is necessary; see also \cite{WeberLe,bmexo,EvansWeberGroup}
for alternative ways to obtain this value.

\begin{figure}[t]
\includegraphics[width=12cm, height=9cm]{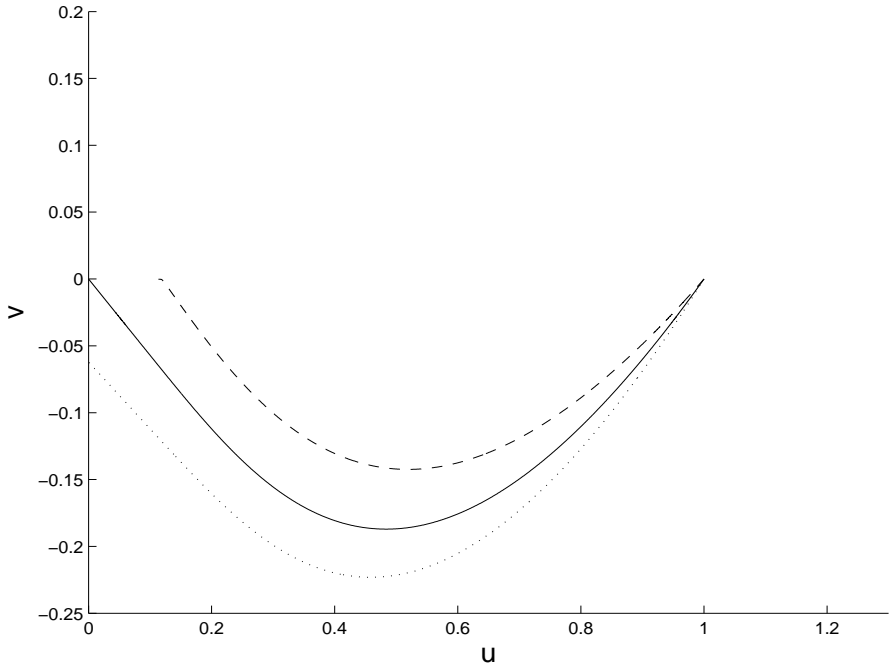}
\caption{Projection onto the $ (u,v) $-plane of trajectories of
(\ref{eq:ode0}) lying on different planes $ H_c = c $.  Here, $ \beta
= 1 $, and the three curves correspond to $ c = 0.5 $ (dotted), $
0.5707 $ (solid) and $ 0.7 $ (dashed). }
\label{fig:findc}
\end{figure}

Thus, we seek a heteroclinic solution of (\ref{eq:ode0}), which
progresses between the fixed points $(1/\beta, 0, 0) $ and $ (0,0,1)
$, and is confined to the plane $ \beta \, v + \beta \, c \, u + c \,
y = c $; i.e., the fuel concentration obeys
\begin{equation}
\label{eq:fuelmass}
y = 1 - \beta \, u - \frac{\beta}{c} \, v
\end{equation}
at all values of $ \xi $.  Considering (\ref{eq:ode0}) under this
restriction, we obtain
\begin{equation}
\label{eq:solid2d}
\left\{
\renewcommand{\arraystretch}{1.3}
\begin{array}{lll}
u^{\prime} & = & \displaystyle v \\
v^{\prime} & = & \displaystyle
- \, c \, v -  \left( 1 - \beta \, u - \frac{\beta}{c} \, v \right) \, e^{-1/u}
\end{array}
\right. \, .
\end{equation}
This is effectively a projection of the flow on the particular
invariant plane $ H_c(u,v,y) $ $ = c $ onto the $ (u,v) $-plane.  Any
value of $ c $ for which a heteroclinic connection exists between $
(1/\beta,0) $ and $ (0,0) $ is a permitted speed for the wavefront.

\begin{figure}[t]
\includegraphics[width=12cm, height=9cm]{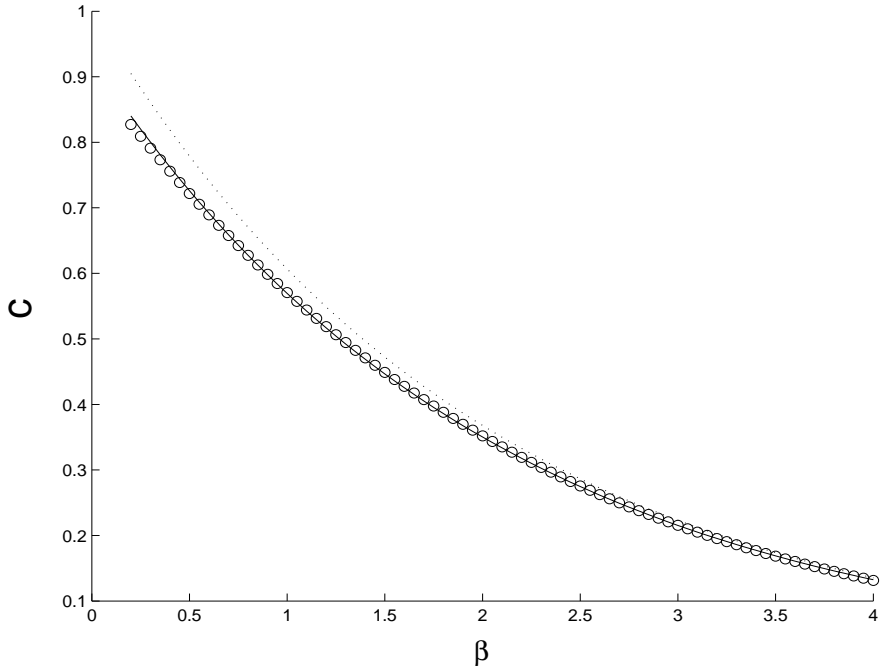}
\caption{Variation of the wavespeed $c$ with $\beta$: open
circles (numerical results); unbroken curve (empirical curve
(\ref{eq:c0})); dotted curve ($ \exp ( -0.5 \beta ) $, as obtained in
\cite{vv,bill,ms}).}
\label{fig:cvsb}
\end{figure}

The unstable eigen-direction of the point $ (1/\beta,0) $ is $ (-c, -
\beta e^{-\beta}) $, and we determine $ c $ numerically by shooting
along this direction, and attempting to match up with a trajectory
approaching the origin.
In Figure~\ref{fig:findc} we show several numerically computed
trajectories of this form, for different values of $ c $, where we
have chosen $ \beta = 1 $.  Note that this is not a standard $ (u,v)
$-phase space for (\ref{eq:solid2d}), since each curve corresponds to
a different value of the parameter $ c $.  Rather, it is a projection
onto the $ (u,v) $-plane of specialized trajectories from the
invariant planes $ H_c(u,v,y) = c $ of the three-dimensional system
(\ref{eq:ode0}).  The one trajectory which makes the required
connection lies in the invariant plane corresponding to $ c = 0.5707
$.  The determination of this $ c $ value was obtained by making
incremental adjustments of $ c $ until an appropriate connection is
obtained.

We use this technique to numerically compute the wavespeeds for
various values of the fuel parameter $\beta$, and illustrate this
dependence in Figure~\ref{fig:cvsb}.  The wavespeed decays with $
\beta $.  For fuels with larger $ \beta $ (poor fuels), the energy
resulting from the reaction is insufficient to quickly activate
combustion in nearby material, and combustion fronts propagate slowly.
The data fits the exponential curve
\begin{equation}
\label{eq:c0}
c(\beta)  = 0.927 \, e^{-0.486 \beta}
\end{equation}
with correlation $ \rho > 0.9999 $.  Equation~(\ref{eq:c0}) therefore
provides an empirically determined formula of excellent accuracy, for
the speed of a wavefront in perfectly solid adiabatic one-dimensional
media.  This result is close (and consistent with) a variety of
sources: $ \exp ( - 0.5 \beta ) $ is quoted in \cite{vv} for the small
$ \beta $ limit; this same value is given as an upper bound in
\cite{bill}, and also implied in eq.~(10) in \cite{ms} using a large $
\beta $ limit within a discontinuous front approximation. See
Figure~\ref{fig:cvsb} for a comparison with our results.

\begin{figure}[t]
\includegraphics[width = 12cm, height = 9cm]{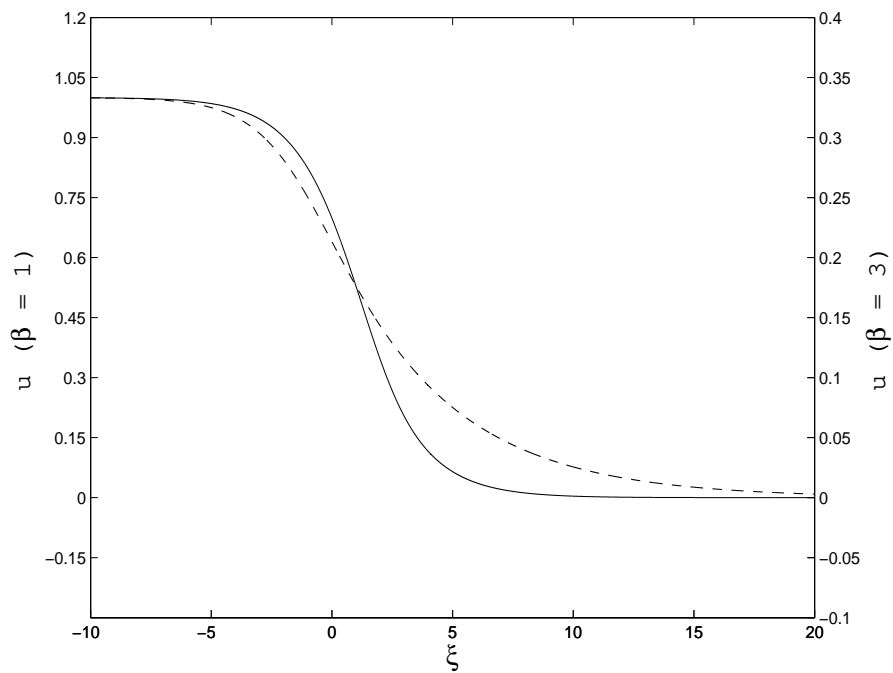}
\caption{Temperature front at $ \beta = 1 $ (solid, left scale) and $ \beta = 3 $ (dashed,
right scale)}
\label{fig:b1andb3}
\end{figure}

The structure of the temperature front is illustrated in
Figure~\ref{fig:b1andb3} for $ \beta = 1 $ (solid curve, left scale)
and $ \beta = 3 $ (dashed curve, right scale), demonstrating that
larger $ \beta $ fronts have a broader preheat layer preceding the
front.  Note that the preheat zone differs from the reaction zone \cite{kapila}.
The latter shrinks with increasing $ \beta $ \cite{bf}.  Specifically,
the reaction zone {\em as a fraction of the preheat zone}
is $ {\mathcal O}(1/\beta) $ \cite{kapila,ms}.  The reaction zone is well
localized near the region of
greatest temperature change \cite{kapila}, and is not immediately
identifiable in temperature profiles as in Figure~\ref{fig:b1andb3}.
Indeed, the increase in size of the preheat layer with $ \beta $
supports the $ {\mathcal O} (1/\beta) $ expectation for the ratio
between the reaction and the preheat zones.

\subsection{Wavespeed for $ 1 \ll \lewis < \infty $}
\label{sec:large}

\begin{figure}[t]
\includegraphics[width = 12cm, height = 9cm]{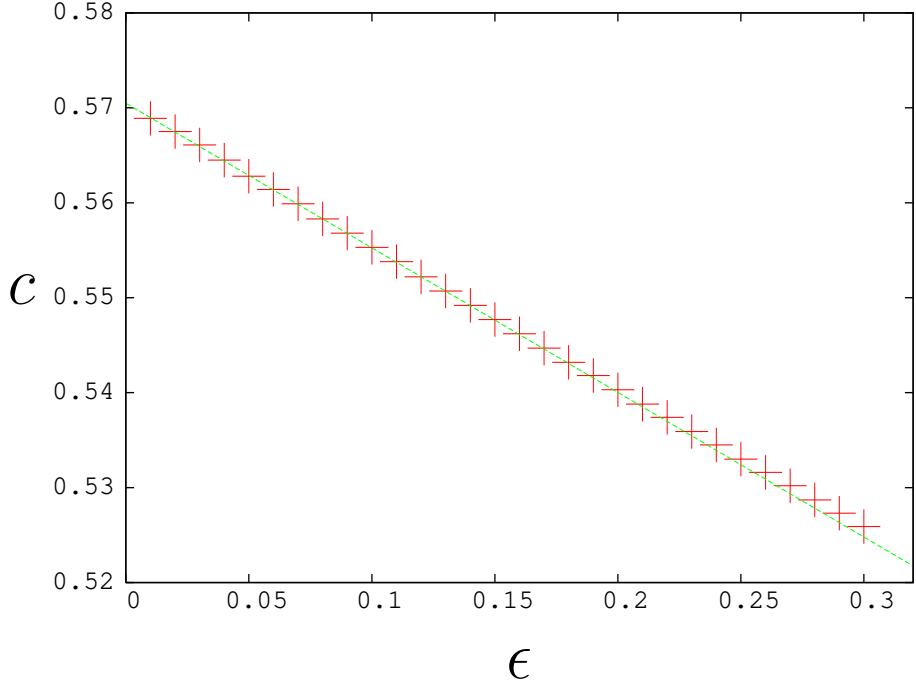}
\caption{Numerically computed wavespeed variation with $ \epsilon $ for $ \beta = 1 $ (the
crosses).  The
dashed line is the theoretical approximation for $ \beta = 1 $ obtained from the
methods of Section~\ref{sec:mel}.}
\label{fig:cvse}
\end{figure}

When the Lewis number is not infinite, but large, $ \epsilon $ is
small, and weak fuel diffusion needs to be permitted.  This is a {\em
singular} limit in (\ref{eq:governpde}) and (\ref{eq:governode}), and
as a consequence has been hardly examined in the literature.  By
defining $ v = u^\prime $ as before, but now also $ z = y^\prime $,
the governing equations (\ref{eq:governode}) can be represented as a
four-dimensional system
\begin{equation}
\label{eq:odeeps}
\left\{
\renewcommand{\arraystretch}{1.3} \begin{array}{lll}
u^\prime & = & \displaystyle v \\
v^\prime & = & \displaystyle - c \, v - y \, e^{-1/u} \\
y^\prime & = & \displaystyle z \\
z^\prime & = & \displaystyle \frac{1}{\epsilon} \left( - c \, z + \beta \, y \, e^{-1/u} \right)
\end{array}
\right. \, .
\end{equation}
This is reducible to three-dimensions: the quantity
\[
G_c^\epsilon(u,v,y,z) = \beta \, v + \beta \, c \, u + c \, y + \epsilon \, z
\]
can be verified to be a conserved quantity of
(\ref{eq:odeeps}). Hence, flow is confined to the invariant
three-dimensional surfaces $ G_c^\epsilon = $ constant.  Now, we seek
a wavefront solution which goes from $ (u,v,y,z) = (u_B,0,0,0) $ to a
value $ (0,0,1,0) $, and we find that $ G_c^\epsilon(u,v,y,z) = c $,
and $ u_B = 1 / \beta $ as before.  The three-dimensional invariant
surface on which both points lie is
\[
z = \frac{1}{\epsilon} \left( c - \beta \, v - \beta \, c \, u - c \, y \right) \, .
\]
The dynamics of (\ref{eq:odeeps}) on this surface can be projected to
the three variables $ (u,v,y) $, such that
\begin{equation}
\label{eq:odeeps3d}
\renewcommand{\arraystretch}{1.3}
\left\{ \begin{array}{lll}
u^\prime & = & \displaystyle v \\
v^\prime & = & \displaystyle - c \, v - y \, e^{-1/u} \\
y^\prime & = & \displaystyle
\frac{1}{\epsilon} \left( c - \beta \, v - \beta \, c \, u - c \, y \right)
\end{array}
\right. \, .
\end{equation}
We seek the value of $ c $ which permits a heteroclinic connection from
$ (u,v,y) = (1/\beta, 0, 0) $ to $ (0,0,1) $.
 The former point (corresponding to $ \xi = - \infty $) has only one
positive eigenvalue, given by $ ( - c + \sqrt{ c^2 + 4 \epsilon \beta e^{-\beta}} ) /
(2 \epsilon) $.
For small $ \epsilon $, we ``shoot'' in the eigen-direction corresponding to this,
with an initial
guess of the wavespeed given by (\ref{eq:c0}).  Thereafter, as in the previous section,
we adjust $ c $ until we obtain a solution which approaches the point $ (0,0,1) $ as
$ \xi \rightarrow \infty $.  We do this numerically by considering the
conditions $ c \, y + \epsilon z = c $
and $ v + c \, u = 0 $ which the front must obey at $ \xi = + \infty $, and using a
root-finding algorithm to adjust $ c $.
For a fixed value $ \beta = 1 $, we illustrate how
the wavespeed $ c $ varies with $ \epsilon $ in Figure~\ref{fig:cvse}, with the crosses.
The dashed curve in Figure~\ref{fig:cvse} is a analytical/numerical approximation we obtain for
the wavespeed in terms of an explicit formula (\ref{eq:crel}).  The next two sections
describe how we obtain this formula.

We notice that $c$ decreases as
we increase $\epsilon$, that is, when we {\em decrease} the Lewis
number.  Now, in dimensional form $ \lewis = \kappa / (\rho \, c_p \,
D) $, where $ \rho, \kappa, c_p $ and $ D $ are the respectively the
density, thermal conductivity, specific heat capacity and molecular
diffusivity of the fuel
\cite{WeberLe,skms,bmexo,bmtravel,2Dand1D}. Increasing $ \epsilon $ is
equivalent to increasing the relative importance of $ D $, $ \rho $
and $ c_p $ in relation to $ \kappa $.  Reducing $ \kappa $ obviously
decreases the ability of heat to move, and hence the combustion
speed. Higher densities result in increased fuel mass in each
location, which means more heat is needed in a given area to ignite
all of the fuel before the wave moves on. Fuels with increased $ c_p $
require more heat to increase the temperature by the a specified
amount. Finally, increasing $ D $ increases the transport of burnt
fuel into the unburnt region and vice-versa, interfering with front
propagation.

We computed the changes to the wavefront profile (akin to
Figure~\ref{fig:b1andb3}) when $ \epsilon $ is changed (not shown).
We verified the obvious physical conclusion that the fuel
concentration front becomes less steep when $ \epsilon $ is increased
from zero.

\subsection{Slow manifold reduction}
\label{sec:reduce}

\begin{figure}[t]
\includegraphics[width = 12cm, height = 6cm]{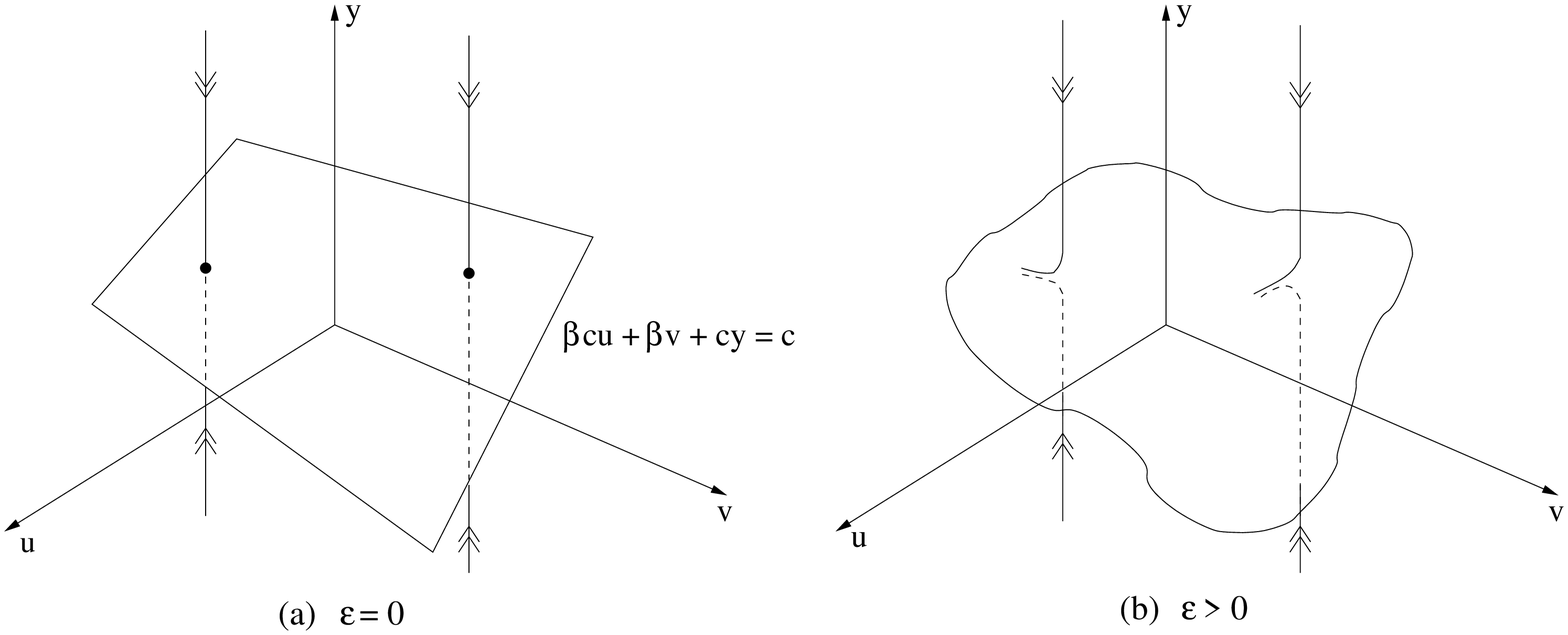}
\caption{The hyperbolic invariant manifold (a) $ {\mathcal S}_0 $ for (\ref{eq:ode3deta}),
and (b) $ {\mathcal S}_\epsilon $ for (\ref{eq:odeeps3deta})}
\label{fig:slowmanifold}
\end{figure}

We now show that in the limit of small $ \epsilon $, it is possible to further
reduce the system (\ref{eq:odeeps3d}) to a {\em two}-dimensional flow on a
{\em slow manifold}.  We begin with
(\ref{eq:odeeps3d}), and note that there are two
``time''-scales in this singularly perturbed system, where we use ``time'' loosely
to mean the independent variable $ \xi $. We therefore
adopt the standard dynamical systems trick of defining a new independent variable $ \eta
=
\xi / \epsilon $ to elucidate motion in the fast ``time'' $ \eta $.
With a dot denoting the rate of change with respect to $ \eta $,
(\ref{eq:odeeps3d}) becomes
\begin{equation}
\label{eq:odeeps3deta}
\left\{ \begin{array}{lll}
\dot{u} & = & \epsilon \, v \\
\dot{v} & = & \epsilon \left( - c \, v - y \, e^{-1/u} \right) \\
\dot{y} & = & c - \beta \, v - \beta \, c \, u - c \, y
\end{array}
\right. \, .
\end{equation}
In the $ \epsilon = 0 $ limit, the system collapses to
\begin{equation}
\label{eq:ode3deta}
\left\{ \begin{array}{lll}
\dot{u} & = & 0 \\
\dot{v} & = & 0 \\
\dot{y} & = & c - \beta \, v - \beta \, c \, u - c \, y
\end{array}
\right. \, ,
\end{equation}
in which it is clear that the plane $ {\mathcal S}_0 $ defined by $ c
- \beta \, v - \beta \, c \, u - c \, y = 0 $ consists entirely of
fixed points.  This is the same plane as defined through $ H_c(u,v,y)
= c $ for equation~(\ref{eq:ode0}), on which the interesting behavior
occurred for perfectly solid fuels.  Each fixed point has a
one-dimensional stable manifold (in the $ y $-direction), and a
two-dimensional center manifold, which is $ {\mathcal S}_0 $.  Thus
the plane $ {\mathcal S}_0 $ is invariant and normally hyperbolic with
respect to (\ref{eq:ode3deta}); there is exponential contraction
towards it as illustrated in Figure~\ref{fig:slowmanifold}(a).

Upon switching on $ \epsilon $ and considering the dynamics
(\ref{eq:odeeps3deta}), $ {\mathcal S}_0 $ perturbs to an invariant
curved entity $ {\mathcal S}_\epsilon $, which is order $ \epsilon $
away from $ {\mathcal S}_0 $.  This is because of the structural
stability of normally hyperbolic sets \cite{fenichel}, which also
implies that normal hyperbolicity is preserved for small $ \epsilon
$. Therefore, there is exponential decay of trajectories towards $
{\mathcal S}_\epsilon $ on ``time''-scales of order $ \eta $, as shown
in Figure~\ref{fig:slowmanifold}(b).  Motion on $ {\mathcal
S}_\epsilon $ occurs at a slower rate (on ``time''-scales of order $
\xi $), and hence it is termed a `slow manifold'.  The heteroclinic
connection we seek lies on $ {\mathcal S}_\epsilon $, from $ (u,v,y) =
(1/\beta, 0,0) $ to $ (0,0,1) $.  Since $ {\mathcal S}_\epsilon $ is
invariant, two-dimensional and not parallel to the $ y $-axis, it
therefore makes sense to project the motion to the $ (u,v) $-plane in
order to describe behavior. To elucidate this motion, we need to once
again return to the original ``time''-scale $ \xi $ -- the slow time
associated with motion on the slow manifold.

Return to the relationship $ G_c^\epsilon \left( u(\xi), v(\xi), y(\xi), z(\xi) \right) = c $,
which upon differentiation yields
\[
\beta \, v^\prime + \beta \, c \, u^\prime + c \, y^\prime + \epsilon \, z^\prime = 0 \, ,
\]
and since $ u^\prime = v $ and $ y^\prime = z $,
\[
z = - \frac{\beta}{c} \, v^\prime - \beta \, v -  \frac{\epsilon}{c}  z^\prime  \, .
\]
Substituting back into $ G_c^\epsilon(u,v,y,z) = c $, we obtain
\[
\beta \, v + \beta \, c \, u + c \, y + \epsilon \left(
- \frac{\beta}{c} \, v^\prime - \beta \, v  + {\mathcal O}(\epsilon) \right) = c \, ,
\]
and thus
\[
y = 1 - \frac{\beta}{c} \, v - \beta \, u + \epsilon \, \frac{\beta}{c^2} \, v^\prime + \epsilon
\, \frac{\beta}{c} \, v +{\mathcal O}(\epsilon^2) \, .
\]
Substitution into the $ v^\prime $ equation in (\ref{eq:odeeps}) or (\ref{eq:odeeps3d}) gives
\[
v^\prime \left( 1 + \epsilon \, \frac{\beta}{c^2} e^{-1/u} \right) =
- c \, v - \left(  1 - \frac{\beta}{c} \, v - \beta \, u  + \epsilon
\, \frac{\beta}{c} \, v +{\mathcal O}(\epsilon^2) \right) e^{-1/u} \, .
\]
Therefore
\begin{eqnarray*}
v^\prime & = & \left( 1 - \epsilon \, \frac{\beta}{c^2} e^{-1/u} \right)
\left[ - c \, v - \left(  1 - \frac{\beta}{c} \, v - \beta \, u  + \epsilon
\, \frac{\beta}{c} \, v  \right) e^{-1/u} \right] +{\mathcal O}(\epsilon^2) \\
& = & - c \, v - \left(  1 - \frac{\beta}{c} \, v - \beta \, u \right) e^{-1/u}
+ \epsilon \, \frac{\beta}{c^2}  \left(  1 - \frac{\beta}{c} \, v - \beta \, u \right)
e^{-2/u} + {\mathcal O}(\epsilon^2) \, .
\end{eqnarray*}
Retaining only $ {\mathcal O}(\epsilon) $ terms, we obtain the $ (u,v) $-projected
approximate equations on the slow manifold
\begin{equation}
\label{eq:odeepsproj}
\renewcommand{\arraystretch}{1.3}
\left\{ \begin{array}{lll}
u^\prime & = & \displaystyle v \\
v^\prime & = & \displaystyle
- c \, v - \left( 1 - \frac{\beta}{c} v - \beta \, u \right) e^{-1/u}
+ \epsilon \, \frac{\beta}{c^2}  \left(  1 - \frac{\beta}{c} \, v - \beta \, u \right)
e^{-2/u}
\end{array} \right. \, .
\end{equation}

We will now show how to use these approximate dynamics to predict the correction
to the wavespeed resulting from the inclusion of the finiteness of the Lewis number.

\subsection{Perturbative formula for wavespeed}
\label{sec:mel}

Here, we derive and numerically study a formula for the wavespeed
correction in going from $ \lewis = \infty $ to finite Lewis number.
Let
\begin{equation}
\label{eq:cexpand}
c \left( \beta ,\epsilon \right)  = c_0 \left( \beta \right) + \epsilon \,
c_1 \left( \beta \right) + {\mathcal O} \left( \epsilon^2 \right) \, ,
\end{equation}
where $ c_0 $ is the wavespeed associated with the infinite Lewis
number ($ \epsilon = 0 $) combustion wavefront.  In the spirit of
perturbation analysis, we obtain a formula for the correction $
c_1(\beta) $ purely in terms of the unperturbed ($\epsilon = 0 $)
wave, using a nontraditional application of ``Melnikov's method''
\cite{melnikov} from dynamical systems theory.

\begin{figure}[t]
\includegraphics[width = 12cm, height = 4cm]{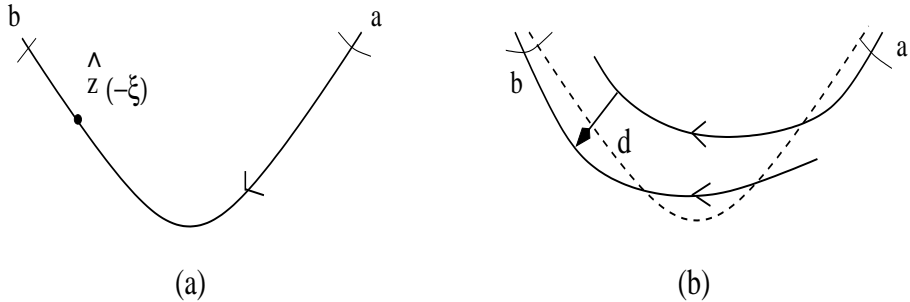}
\caption{Manifold structure for the Melnikov approach: (a) $ \epsilon = 0 $,
(b) $ \epsilon \ne 0 $}
\label{fig:mel}
\end{figure}

Melnikov's method is applied most commonly to area-preserving
two-dimensional systems under time-periodic perturbations
\cite{gh,ap,wiggins}. (Here, once again, $ \xi $ represents ``time''.)
Our system (\ref{eq:odeepsproj}) is not area-preserving, and has a
perturbation which is independent of the temporal variable.  Under
these conditions, we describe the method applied to the system
\begin{equation}
\label{eq:melset}
{\mathbf z}^\prime = {\mathbf f} \left( {\mathbf z} \right) + \epsilon \,
{\mathbf g} \left( {\mathbf z} \right) \, .
\end{equation}
When $ \epsilon = 0 $, suppose this system possesses a heteroclinic
connection between the two saddle fixed points $ {\mathbf a} $ and $
{\mathbf b} $ as shown in Figure~\ref{fig:mel}(a).  A heteroclinic
connection of this sort occurs when a branch of the one-dimensional
unstable manifold of $ {\mathbf a} $ coincides with a branch of the
stable manifold of $ {\mathbf b} $.  This heteroclinic trajectory can
be represented as a solution $ {\mathbf z} =
\hat{\mathbf z}(\xi) $ to (\ref{eq:melset}) with $ \epsilon = 0 $.

Now, for small $ \epsilon > 0 $ in (\ref{eq:melset}), the fixed points
$ {\mathbf a } $ and $ {\mathbf b} $ perturb by $ {\mathcal
O}(\epsilon) $, and retain their stable and unstable manifolds
\cite{fenichel}.  However, these need no longer coincide.
Figure~\ref{fig:mel}(b) shows how they can split apart, with the
dashed curve showing the original manifold.  Let $ d(\xi,\epsilon) $
be a distance measure between these manifolds, measured along a
perpendicular to the unperturbed heteroclinic drawn at $ \hat{\mathbf
z}(-\xi) $.  The variable $ \xi $ can thus be used to identify the
position along the heteroclinic curve (cf.\ ``heteroclinic
coordinates'' of Section~4.5 in \cite{wiggins}).  Since $ d(\xi,0) = 0
$ for all $ \xi $, this distance is Taylor expandable in $ \epsilon $
in the form
\[
d(\xi,\epsilon) = \epsilon \, \frac{M(\xi)}{ \left| {\mathbf f} \left( \hat{\mathbf z}(-\xi)
\right) \right|} + {\mathcal O}(\epsilon^2) \, ,
\]
where the scaling factor $ \left| f \left( \hat{\mathbf z}(-\xi)
\right) \right| $ in the denominator represents the unperturbed
trajectory's speed at the location $ \xi $.  The quantity $ M(\xi) $
is the ``Melnikov function'', for which an expression is
\begin{equation}
\label{eq:melformula}
M \left( \xi \right) = \int_{-\infty}^{\infty} \exp\left[-\int_{-
\xi}^{\mu} {\mathbf \nabla}
\cdot {\mathbf f}(\hat{\mathbf z}(s)) \, ds\right] {\mathbf f} (\hat{\mathbf z}(\mu))
\wedge {\mathbf g}(\hat{\mathbf z}(\mu)) \, d\mu \, ,
\end{equation}
where the wedge product between two vectors is defined by $
(a_1,a_2)^T \wedge (b_1,b_2)^T = a_1 b_2 - a_2 b_1 $.  Obtaining the
version (\ref{eq:melformula}) requires two adjustments to the standard
Melnikov approaches \cite{gh,ap,wiggins}: incorporating the non
area-preserving nature of the unperturbed flow of (\ref{eq:melset}),
and representing the distance in terms of heteroclinic coordinates.
Details are provided in Appendix~\ref{app:mel}.  We need to ensure the
persistence of a heteroclinic trajectory in (\ref{eq:odeepsproj}) for
$ \epsilon > 0 $, and thus require $ d(\xi,\epsilon) = 0 $ for all $
\xi $.  For this to happen for all small $ \epsilon $, we therefore
need to set $ M(\xi) \equiv 0 $.

To apply this technique to our system, we begin by writing
(\ref{eq:odeepsproj}) in the form (\ref{eq:melset}).  Using the
expansion (\ref{eq:cexpand}), and utilizing binomial expansions for $
1 / (c_0 + \epsilon c_1) $, we get
\begin{equation}
\label{eq:epsexpand}
\left\{ \begin{array}{lll}
\renewcommand{\arraystretch}{1.3}
\displaystyle u^\prime & = & \displaystyle v \\
\displaystyle v^\prime & = & \displaystyle - c_0 \, v - e^{-1/u} \Upsilon_{uv} + \epsilon \left(
- c_1 v - \frac{\beta c_1 e^{-1/u}}{c_0^2} v + \frac{\beta e^{-2/u}}{c_0^2} \Upsilon_{uv} \right)
\end{array} \right. \, ,
\end{equation}
where higher-order terms in $ \epsilon $ have been discarded, and
\[
\Upsilon_{uv} = 1 - \beta \, u - \frac{\beta}{c_0} \, v \, .
\]
By appropriately identifying $ {\mathbf f} $ and $ {\mathbf g} $ from
(\ref{eq:epsexpand}) through comparison with (\ref{eq:melset}), we see
that
\[
\left( {\mathbf f} \wedge {\mathbf g} \right) (u,v) =
v \left( - c_1 v - \frac{\beta c_1 e^{-1/u} v}{c_0^2}
+ \frac{\beta e^{-2/u}}{c_0^2} \Upsilon_{uv} \right)
\]
and $ {\mathbf \nabla} \cdot {\mathbf f} = - c_0 + \beta e^{-1/u} /
c_0 $. Substituting into the Melnikov formula (\ref{eq:melformula}),
and setting it equal to zero, we obtain
\[
\int_{-\infty}^{\infty}\exp \left[ \int_{-\xi}^{\mu} \left( c_0 - \frac{\beta}{c_0}e^{-1/u(s)}
 \right) ds \right] v \left(-c_1v - \frac{\beta v
c_1 e^{-1/u}}{c_0^2} + \frac{\beta e^{-2/u}}{c_0^2} \Upsilon_{uv} \right) \, d\mu = 0 \, ,
\]
where each of $ u(\mu) $ and $ v(\mu) $ is evaluated along the $
\epsilon = 0 $ combustion wave.  Notice, however, that for this
infinite Lewis number combustion wave, (\ref{eq:fuelmass}) tells us
that the fuel concentration $ y(\mu) = \Upsilon_{uv}(\mu) $ for all $
\mu $.  Therefore
\begin{equation}
\label{eq:c1}
c_1 (\beta) = \beta \, \frac{\displaystyle \int_{-\infty}^{\infty}\exp\left[\int_{-\xi}^{\mu}
\left( c_0 - \frac{\beta}{c_0} e^{-1/u(s)} \right) \, ds\right] v \, y \, e^{-2/u} \,
d\mu}{\displaystyle \int_{-\infty}^{\infty}\exp\left[\int_{-\xi}^{\mu} \left( c_0
- \frac{\beta}{c_0} e^{-1/u(s)} \right) \, ds
\right] v^2 (c_0^2 + \beta \, e^{-1/u}) \, d\mu} \, ,
\end{equation}
where $ u(\mu) $, $ v(\mu) $ and $ y(\mu) $ in the integrands are
obtained from the $ \epsilon = 0 $ combustion wave discussed in
Section~\ref{sec:infty}.  The apparent dependence of $ c_1 $ on the
wave coordinate $ \xi $ is spurious: if $ I $ is an anti-derivative of
the inner integrals in (\ref{eq:c1}), a multiplicative term $ \exp
\left[ - I(-\xi) \right] $ emerges in both the numerator and
denominator, which therefore cancels.  Hence, any convenient value for
$ \xi $ can be chosen in (\ref{eq:c1}), for example $ 0 $.

Equation~(\ref{eq:c1}) is a powerful expression in which the wavespeed correction
is expressed purely in terms of the (unperturbed) infinite Lewis number wavefront
and system parameters.  This
correction was obtained through an application of the slow manifold and Melnikov's method
(suitably modified).  While developed within the current specific context, we note that
this technique can in fact be used in a variety of instances which are modeled through
coupled reaction-diffusion equations in which the diffusivities are very different.

\begin{figure}[t]
\includegraphics[width = 12cm, height = 8cm]{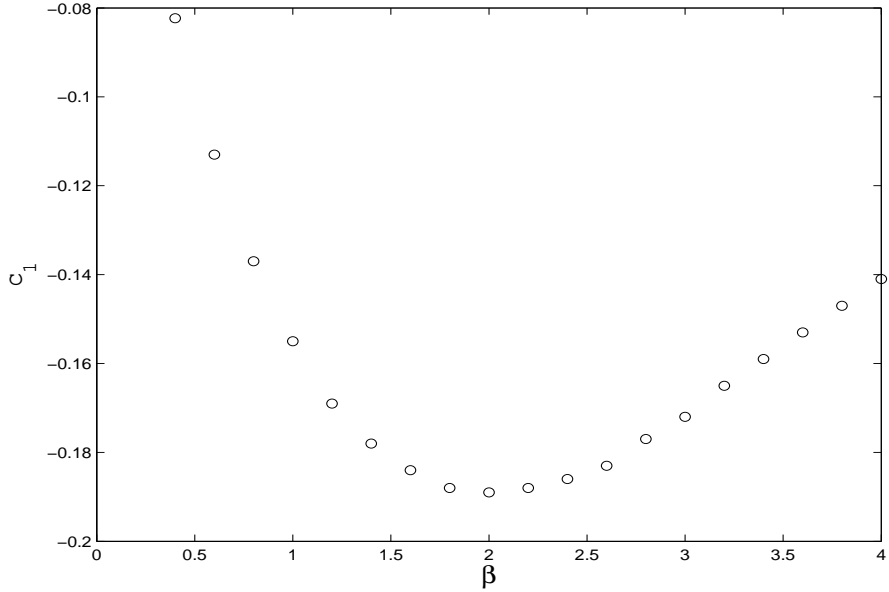}
\caption{The perturbing wavespeed as a function of $ \beta $}
\label{fig:c1vsb}
\end{figure}

We note that $ v < 0 $ for the infinite Lewis number wavefront, as is
clear from the phase-portrait Figure~\ref{fig:findc}. Alternatively, $
u $ is smaller at the front of the wave, where fuel is yet to be
burnt, and is therefore a decreasing function of $ \mu $, leading to $
v = u^\prime < 0 $.  Based on this, (\ref{eq:c1}) immediately displays
that $ c_1 < 0 $, proving the property that the wavespeed decreases
when fuel diffusivity is included.  This is in agreement with the
numerical observations in Section~\ref{sec:large}.

Equation~(\ref{eq:c1}) provides an explicit perturbative formula on
how the wavespeed varies through the inclusion of the finiteness of
the Lewis number, expressed entirely in terms of the infinite Lewis
number combustion wave.  This result is used to compute the solid line
in Figure~\ref{fig:cvse}, which is the theoretical wavespeed $ 0.5707
- 0.1552 \epsilon $ obtained by using (\ref{eq:c1}) and
(\ref{eq:cexpand}) when $ \beta = 1 $.  When $ \epsilon $ is small, it
forms an excellent approximation to the numerically obtained
wavespeed, as described in Section~\ref{sec:large}.  Indeed,
Figure~\ref{fig:cvse} show that the theoretical line is tangential to
the curve formed by the closed circles.

\begin{figure}[t]
\includegraphics[width = 12cm, height = 8cm]{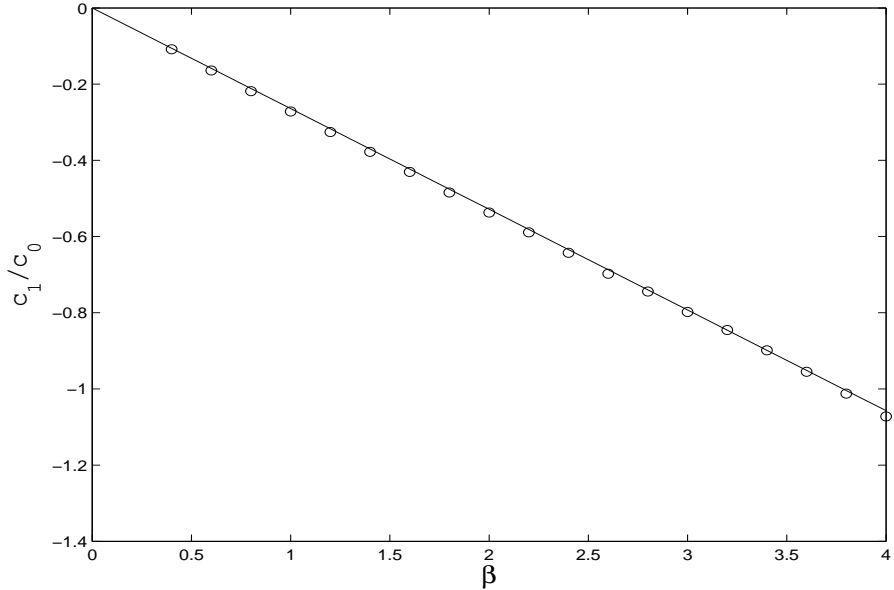}
\caption{Relative size of perturbative wavespeed as a function of $ \beta $}
\label{fig:crelvsb}
\end{figure}

The perturbation wavespeed $ c_1 $ as a function of $ \beta $ appears
in Figure~\ref{fig:c1vsb}.  There is a value of $ \beta $ (around $ 2
$) at which the absolute influence of the finiteness of the Lewis
number is greatest. Nevertheless, since $ c_0 $ is itself a function
of $
\beta $, it makes sense to investigate the {\em relative} influence
$ c_1 / c_0 $ of the perturbative term.  Such is presented in the
numerically computed Figure~\ref{fig:crelvsb}.  The graph is virtually
linear and has zero intercept.  In other words, the complicated
quotient in (\ref{eq:c1}) is in fact proportional to the unperturbed wavespeed $ c_0(\beta) $,
with the proportionality factor {\em independent of $ \beta $}.
We therefore arrive at the approximation
\begin{equation}
c(\beta,\epsilon) = c_0(\beta) \left[ 1 - 0.267 \, \epsilon \, \beta \right] = c_0(\beta)
\left[ 1 - 0.267 \, \frac{\beta}{\lewis} \right] \, ,
\label{eq:crel}
\end{equation}
for large Lewis numbers, with excellent validity across all $ \beta $,
and with $ c_0(\beta) $ also known through (\ref{eq:c0}).

Equation~(\ref{eq:crel}) shows that the wavespeed, as a
fraction of the infinite Lewis number wavespeed, acquires a correction linear in
the ratio $ \beta / \lewis $.  We are not aware of any such result being previously
reported in the literature of combustion waves.  Moreover, the simplicity of this
expression is remarkable.
For the specific instance $ \beta = 1 $, we apply this formula in order to arrive
at the dashed line in Figure~\ref{fig:cvse}.  Our perturbative
theory has clearly given us a very accurate and simple approximation, and elucidates
the straightforward dependence of the wavespeed on the parameters $ \beta $ and
$ \lewis $.

\section{Stability analysis}
\label{sec:stability}

In this section, we test the stability of the combustion wavefront
$ (u,y) =  \left( u_0(\xi),y_0(\xi) \right) $ we have found as a solution to
(\ref{eq:governpde}) at large Lewis numbers.  Consider a perturbation of the form
\begin{equation}
u=u_0(\xi)+U(\xi)\,e^{\lambda\,t},\;\;y=y_0(\xi)+Y(\xi)\,e^{\lambda\,t}.
\label{exppert}
\end{equation}
At first order, $U$ and $Y$ satisfy an eigenvalue problem
\begin{equation}
\left(
\begin{array}{l}
U\\V\\Y\\Z
\end{array}
\right)'=
\left(
\begin{array}{ccccccc}
0&&1&&0&&0\\
\lambda-\frac{y_0}{u_0^2}\,e^{-1/u_0}&&-c&&-e^{-1/u_0}&&0\\
0&&0&&0&&1\\
\frac{\beta\,y_0}{\epsilon\,u_0^2}\,e^{-1/u_0}&&0&&\frac{\lambda}{\epsilon}+
\frac{\beta}{\epsilon}\,e^{-1/u_0}&&-\frac{c}{\epsilon}
\end{array}
\right)
\left(
\begin{array}{l}
U\\V\\Y\\Z
\end{array}
\right).
\label{linearinf2}
\end{equation}

Linear instability occurs if there are values of $ \lambda $ in the
right half plane for which (\ref{linearinf2}) possesses a solution uniformly bounded
for all $ \xi $.
It turns out that such values of $ \lambda $ can be investigated by analyzing the
Evans function \cite{Ev77}, which is a complex analytic function $ E(\lambda) $ whose
zeros correspond to exactly these $ \lambda $ values.  If for example it can be shown
that $ E(\lambda) $ has no zeros in the right half plane, the indications from the
point spectrum of (\ref{linearinf2}) is that the wavefront is stable.  If there exist
zeros of $ E(\lambda) $ in the right half plane, the wavefront is unstable.  A description
of the Evans function as used in our study is given in Appendix~\ref{app:evans}.
This was proposed in \cite{Al90,Ev75,Jo84,Ya85} and has been used by \cite{EvansWeberGroup} for a detailed numerical analysis of (\ref{eq:governpde}).
(It must
be also mentioned that in the linear stability analysis, it is necessary to consider the
essential spectrum associated with (\ref{linearinf2}); it turns out that this has no
intersection with the right half plane and therefore need not worry us further.)

We begin with a traveling wave solution $u_0(\xi)$ and $y_0(\xi)$ obtained using
standard shooting methods in Section~\ref{sec:wavespeed}, and
then compute the Evans function using the procedure outlined in Appendix~\ref{app:evans}.
We note that the system is very sensitive due to its
stiffness. We found a solution to be accurate enough if we obtain
$E(\lambda=0)\sim{\cal{O}}(10^{-12})$. We are guided in our calculations
by the detailed stability analysis of Gubernov {\em et al} \cite{EvansWeberGroup}.  They show, for
example, the lack of any eigenvalues of positive real part for small
$ \beta $, but show that for $ \beta $ large enough, two eigenvalues
pop into the right half plane exhibiting a Hopf bifurcation.  Physically,
this corresponds to a pulsating instability in the wavefront, a well
known phenomenon also occurring for $ \lewis = \infty $ even in slightly different models \cite{WeberLe,Cardarelli,ms,bmtwo}.
Gubernov {\em et al} extend these infinite Lewis number analyses by producing in Figure~5 in \cite{EvansWeberGroup} the stability
boundary in $ \beta $-$ \epsilon $ space (their $ \tau $ is our $ \epsilon $).
We verify here that our numerically computed wavefronts display the characteristics
outlined by them.

\begin{figure}[t]
\includegraphics[width = 12cm, height = 9cm]{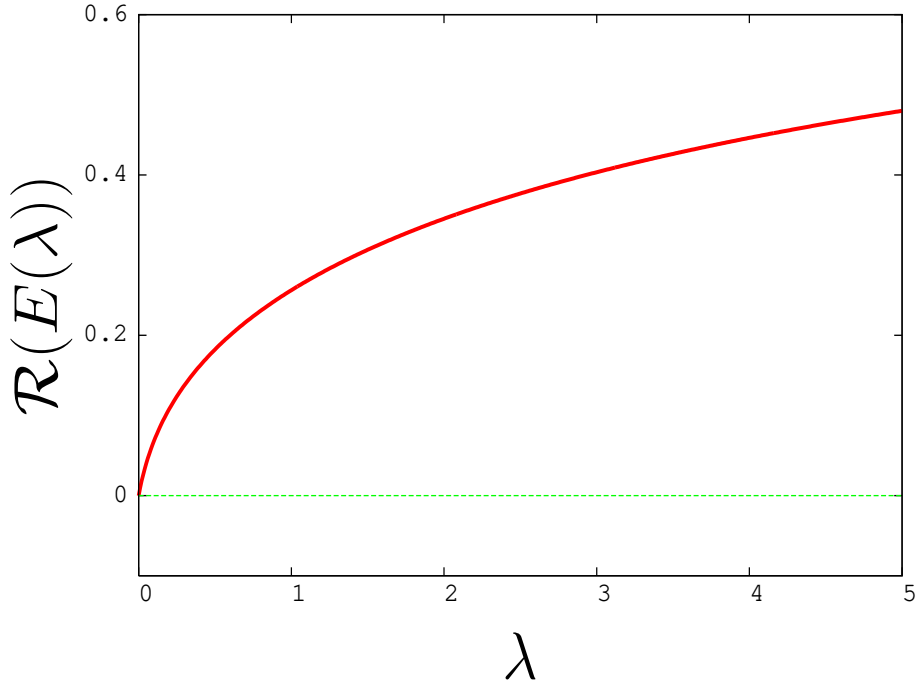}
\caption{The real part of the Evans function $E(\lambda)$ for $\lewis=17$ and
$\beta=1$ as a function of $\lambda$.}
\label{fig:E1}
\end{figure}

\begin{figure}[t]
\includegraphics[width = 6cm, height = 4.5cm]{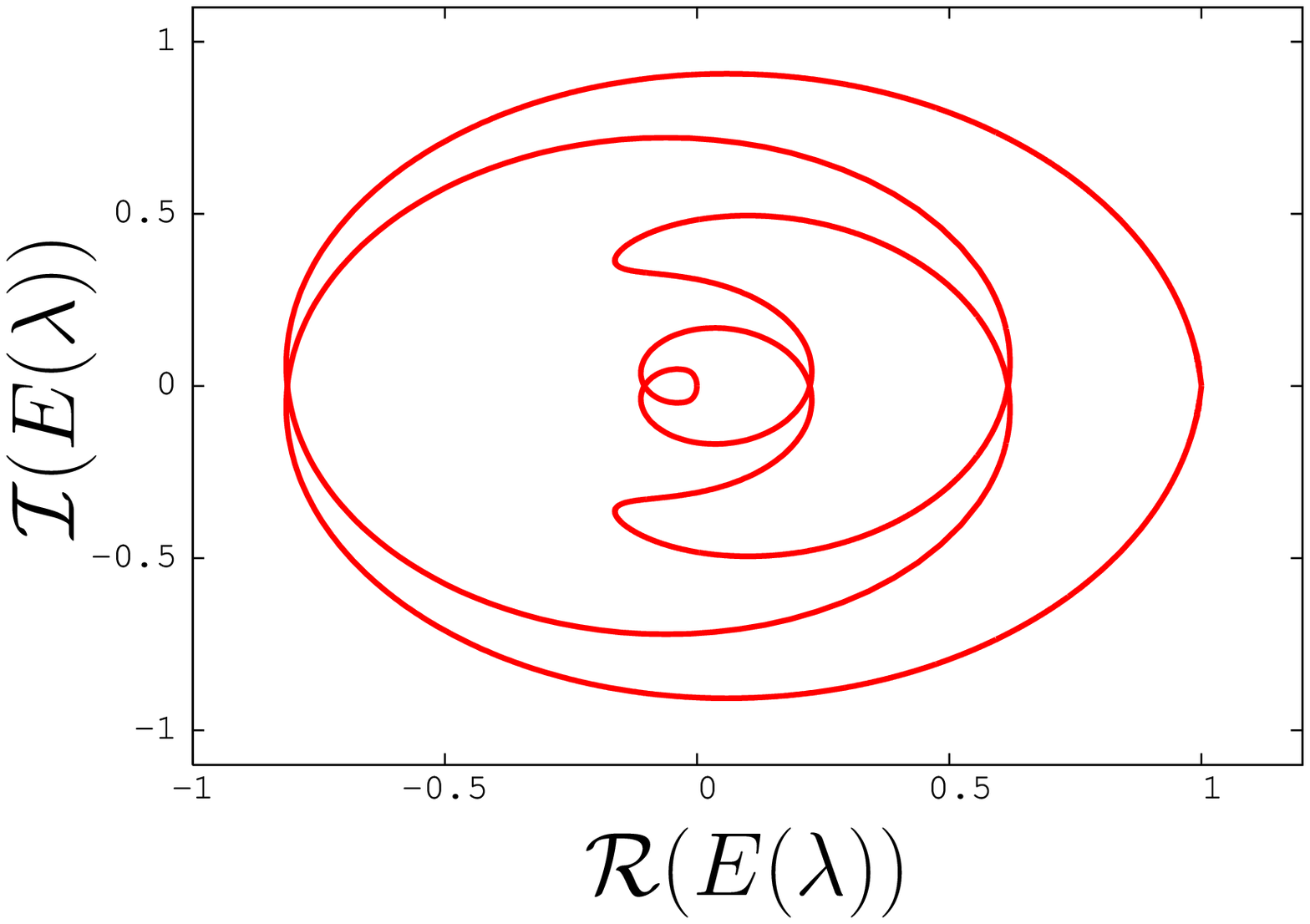}
\includegraphics[width = 6cm, height = 4.5cm]{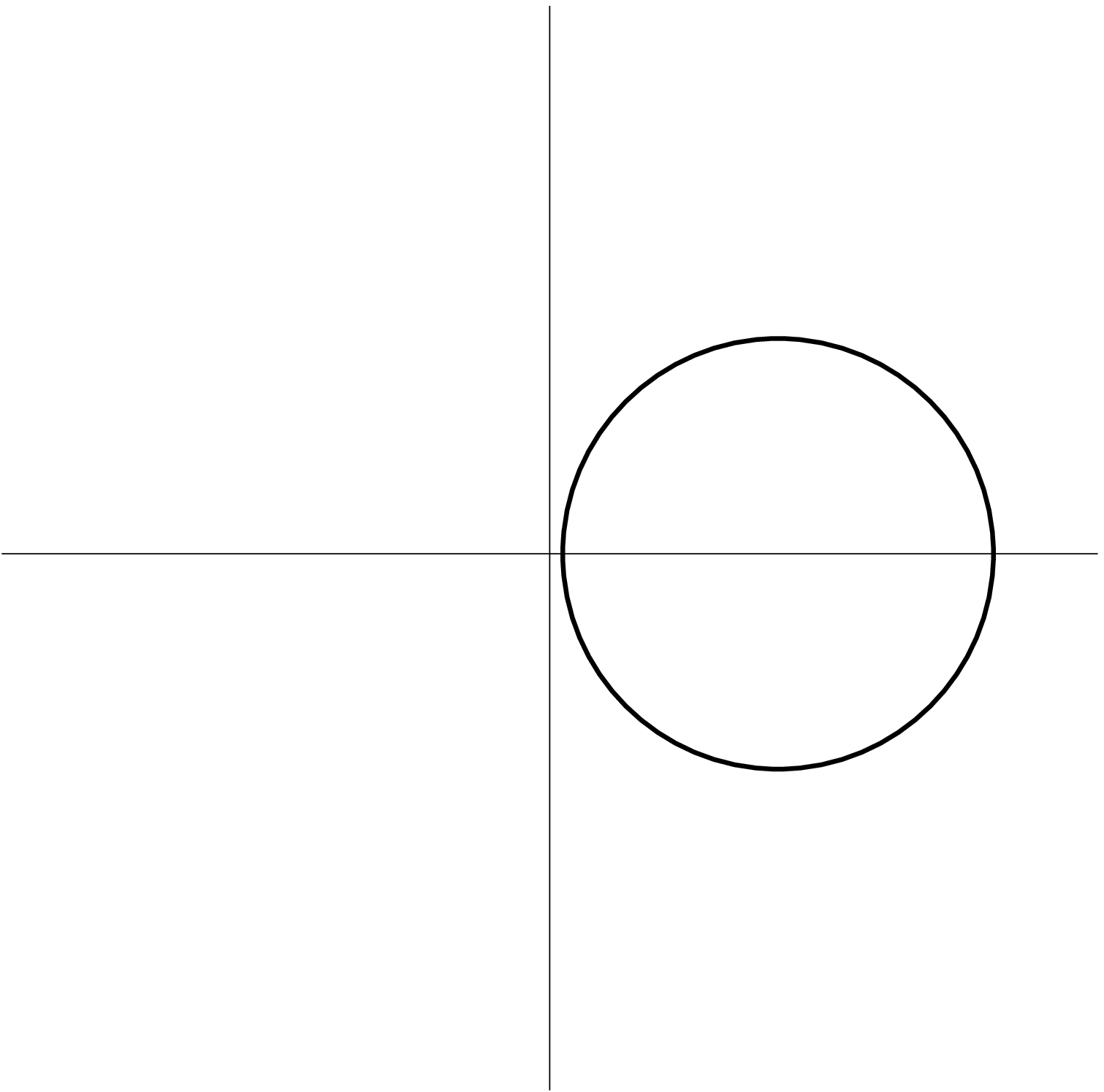}
\caption{Left: The real versus the imaginary part of the Evans function
$E(\lambda)$ for $\lewis=17$ and $\beta=1$. The spectral parameter
$\lambda$ varies along the imaginary axis. Right: A sketch of a
topologically equivalent Evans function. The winding number is clearly
zero indicating stability.}
\label{fig:E23}
\end{figure}

In Figure~\ref{fig:E1} we show the Evans function $E(\lambda)$ as it
varies with increasing $\lambda\in \R$ for $ \lewis=17$ and $\beta=1$ (this
corresponds to a stable regime in Figure~5 of \cite{EvansWeberGroup}). We
find that Evans function does not have any positive real roots. To test for
complex roots we vary $\lambda \in i\R$; using Cauchy's theorem
we can calculate the winding number to detect possible oscillatory
instabilities. In the left panel of Figure~\ref{fig:E23} we show the
complex Evans function. Since the system (\ref{eq:governpde}) is
translationally invariant, the Evans function has at least a simple
zero at $\lambda=0$.  We checked with a little off-set of the order
${\cal{O}}(10^{-5})$ whether the (real) value of the Evans function at
$\lambda=0$ is shifting towards larger values or smaller values. The
off-set allows us to integrate parallel to the imaginary axis of
$\lambda$ and therefore excluding the zero of the Evans function
stemming from the root at $\lambda=0$. This enables us to attribute
roots of the Evans function to either the translational mode or to a
real instability. For the case depicted in Figure~\ref{fig:E23} we find
that the Evans function moves to the right. This means that the Evans
function can be cast in the topologically equivalent form depicted in
the right panel of Figure~\ref{fig:E23} and it has clearly a winding
number zero. We therefore find that at these
parameter values there are no unstable eigenvalues. (Note that for this argument to work
we need our definition
of the Evans function to be analytic which excludes standard methods
such as Gram-Schmidt orthogonalizations.)

\begin{figure}[t]
\includegraphics[width = 12cm, height = 9cm]{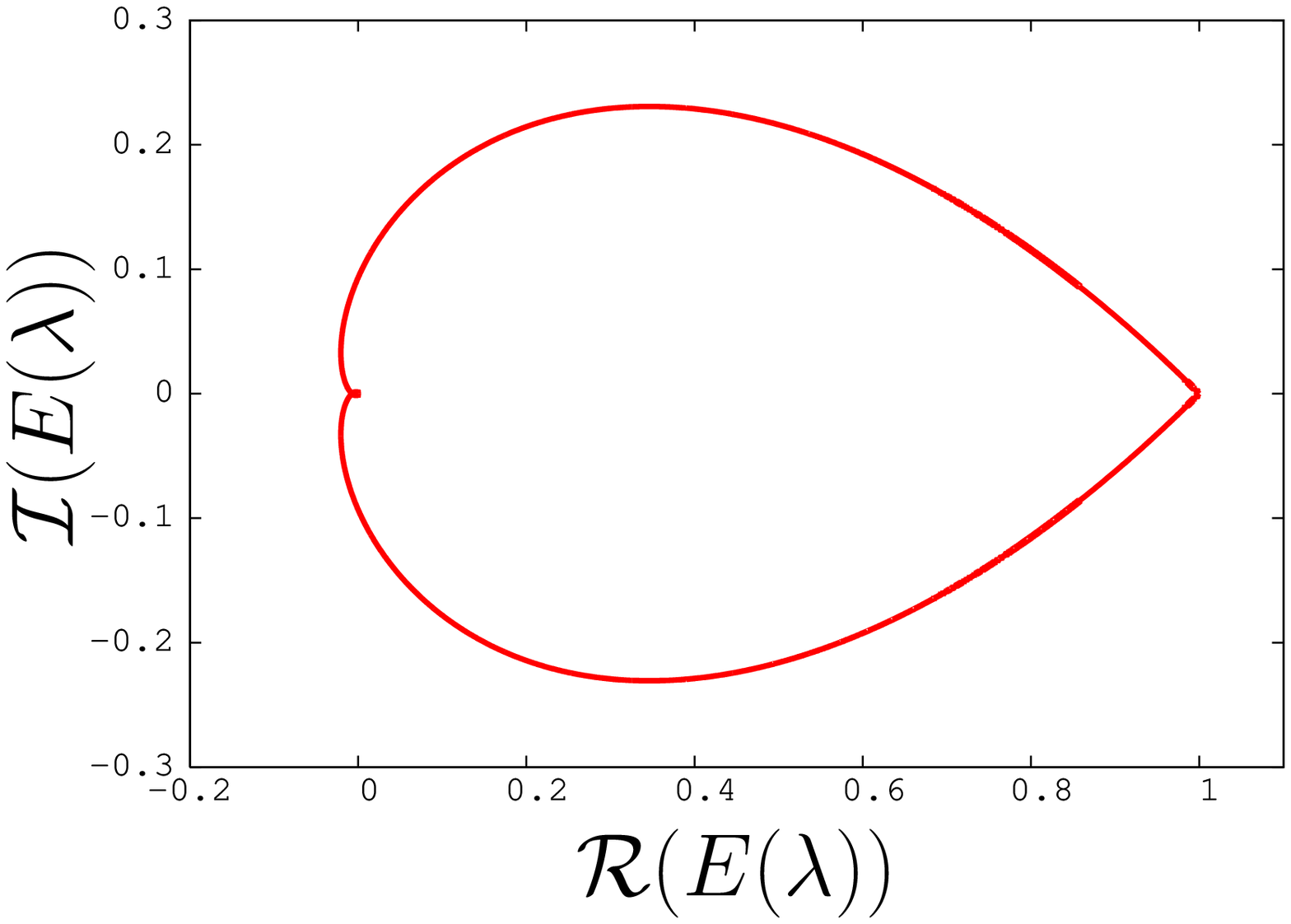}
\caption{The real versus the imaginary part of the Evans function
$E(\lambda)$ for $\lewis=100$ and $\beta=9$. The spectral parameter
$\lambda$ varies along the imaginary axis.}
\label{fig:E4}
\end{figure}

\begin{figure}[t]
\includegraphics[width = 6cm, height = 4.5cm]{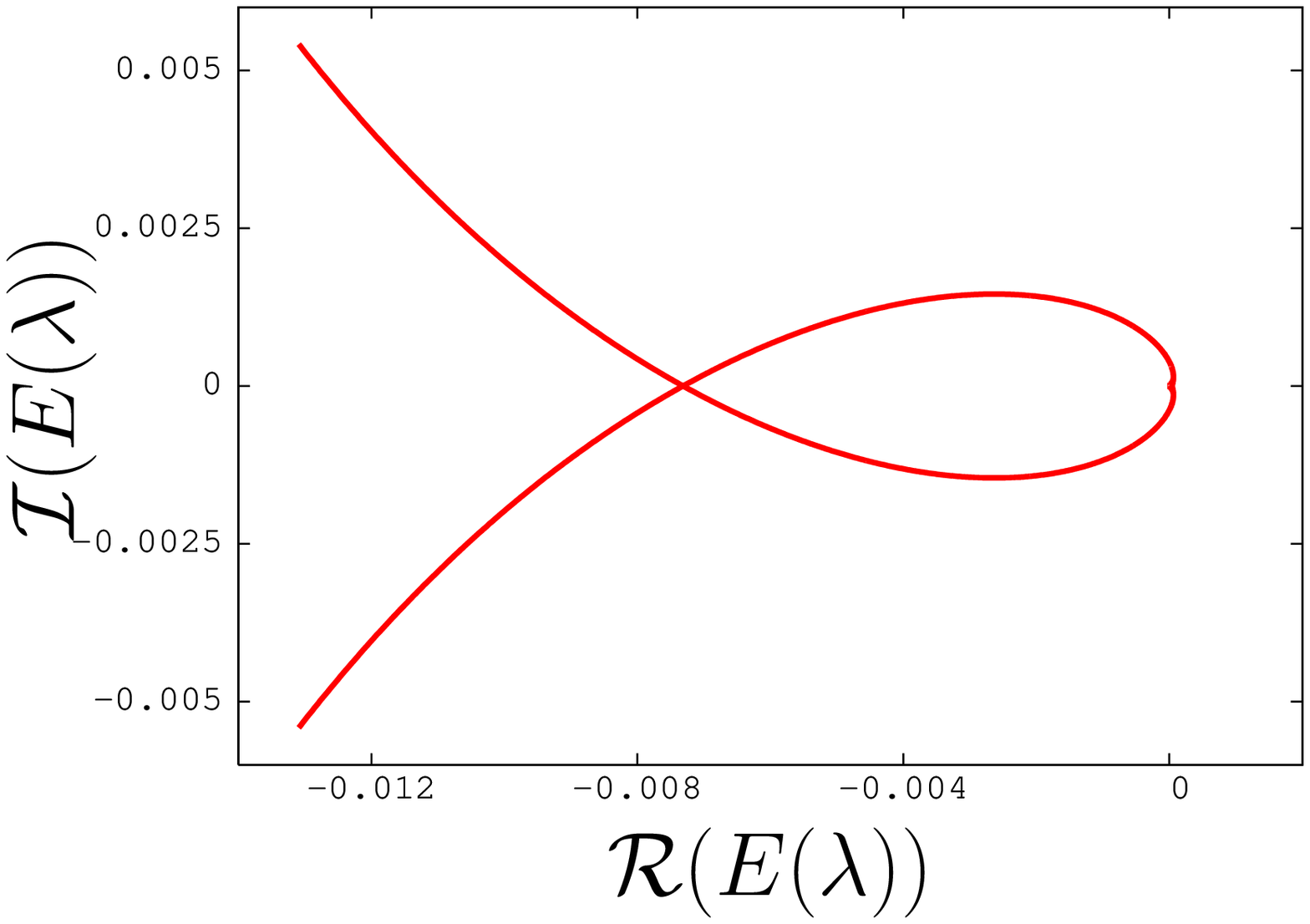}
\includegraphics[width = 6cm, height = 4.5cm]{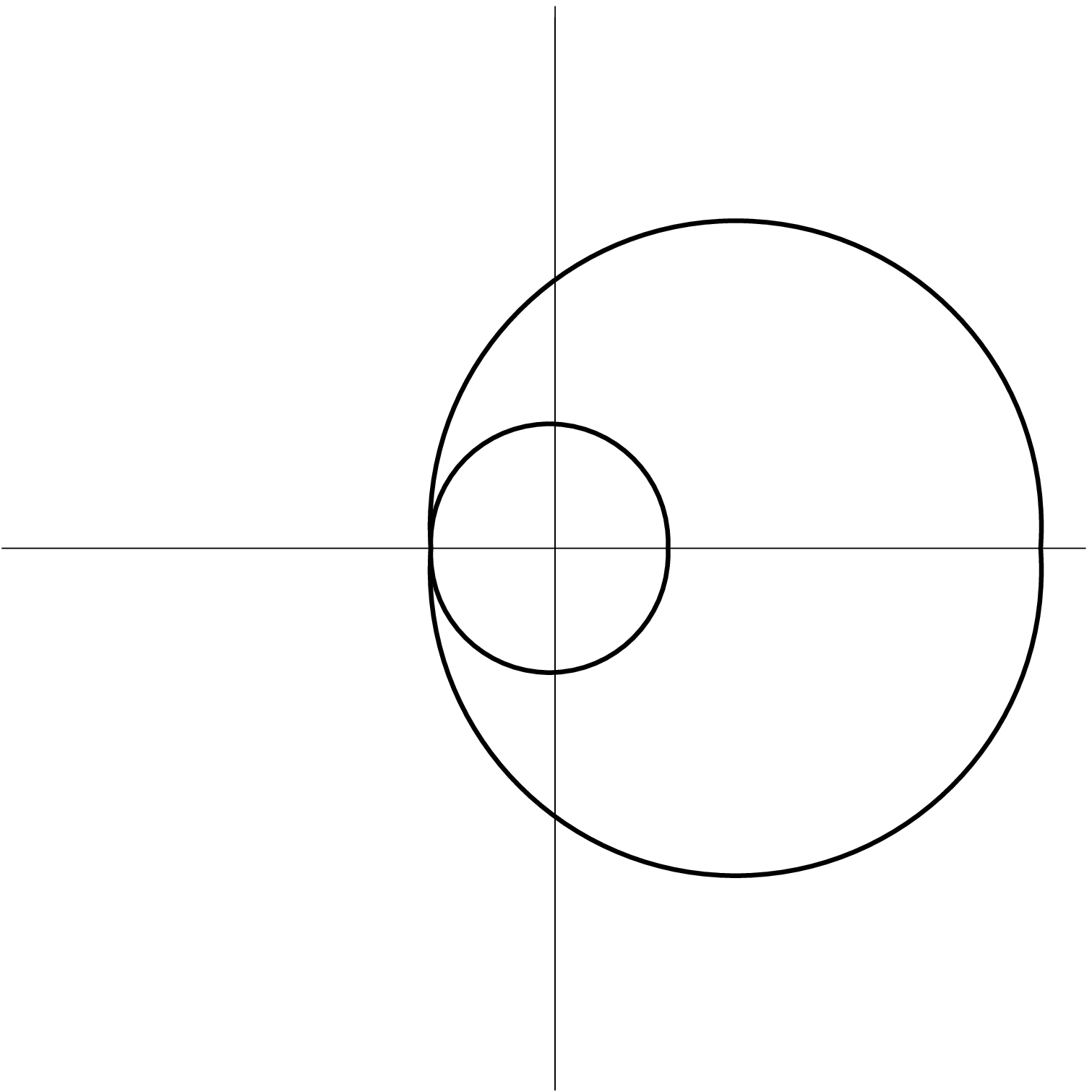}
\caption{Left: Closeup of the Evans function depicted in
Fig.~\ref{fig:E4} into the region $E(\lambda)=0$. Right: A sketch of a
topologically equivalent Evans function. The winding number is clearly
two indicating an oscillatory instability.}
\label{fig:E56}
\end{figure}

We next choose $ \lewis = 100 $ and $\beta = 9$, parameters at which
(according to Figure~5 in \cite{EvansWeberGroup}) an oscillatory instability
is to be expected. In
Figure~\ref{fig:E4} we show the Evans function in this situation, with a
zoom in displayed in Figure~\ref{fig:E56}. To determine the winding number we
need to check whether the small circle of the Evans function in
Figure~\ref{fig:E56} includes the zero or not. We can do so by
allowing again for a small off-set of $\lambda$. We find that the
circle includes zero. Unfolding the behavior of the Evans function
then allows to sketch a topologically equivalent Evans function as in
the right panel of Figure~\ref{fig:E56}. We verified that the
instability is indeed oscillatory by examining the
Evans function for $\lambda \in \R^+$, which reveals no zeros. Hence
our wavefront displays the predicted characteristics of \cite{EvansWeberGroup}.
Stability properties are not affected unduly by the finiteness of the Lewis
number, despite the singularity of this limit.

\section{Concluding remarks}
\label{sec:Conclusion}

In this article, we have studied combustion wavefront in a one-dimensional medium.
Our concentration
was on very high Lewis numbers relevant to high-density
supercritical combustion.  We determine the wavespeed as a function of
the exothermicity parameter $ \beta $ and the Lewis number $ \lewis$,
by seeking the wavespeed value which
establishes a connection between the fixed points corresponding to the
fully burnt and the unburnt states.  The infinite Lewis number instance
reveals an exponential dependence of the wavespeed on $ \beta $, for which
we determine an empirical formula.  We then use several suitably modified
dynamical systems techniques (slow manifold reduction, and Melnikov's method)
to compute an explicit
formula (\ref{eq:c1}) for the correction to the wavespeed when including the effect of
large, but not infinite, Lewis number.  We hence obtain a simple formula
(\ref{eq:crel}) which shows that the relative change in the wavespeed is proportional
to $ \beta / \lewis $ for large Lewis numbers.  Our theory is shown to have
excellent consistency with the numerically computed wavespeed for large
$ \lewis $, as we show in Figure~\ref{fig:cvse}.

The stability of the high Lewis number wavefronts is then
tested numerically based on the Evans function technique.   Our results are
in agreement with
the stability boundaries presented by Gubernov {\em et al} in Figure~5 in
\cite{EvansWeberGroup}.

We remark that the modified Melnikov's method that we have used can in fact be
used in more general situations which are described by two coupled
reaction-diffusion equations with strongly differing diffusivities.  Based on a
known wavefront or wave-pulse solution for when the smaller of the diffusivities
is zero, our technique can in some instances be used to determine the wavespeed correction
resulting from the inclusion of the (previously neglected) diffusivity.  Alternatively,
it can be adapted to situations in which the wavespeed changes due to some
other small parameter.  Our analysis
of high Lewis number wavefronts therefore provides a new perturbative methodology for
analyzing certain classes of reaction-diffusion equations, pattern formation problems
and combustion waves.

{\bf Acknowledgements} \\ The authors wish to thank Harvinder Sidhu,
Konstantina Trivisa, Marshall Slemrod, and Joceline Lega for
discussions and pointers.  Detailed comments and suggestions from two
anonymous referees, whose efforts revealed a substantial error in a
previous version of this manuscript, are also gratefully acknowledged.
Part of this work was done while S.L.\ was
visiting the School of Mathematics and Statistics at the University of
Sydney and while S.B.\ was visiting the Department of Mathematics at
the College of Charleston.  This material is
based upon work supported by the National Science Foundation under
Grant No. DMS-0509622 to S.L. G.A.G.\ gratefully acknowledges support by
the Australian Research Council, DP0452147.

\appendix

\section{Melnikov function derivation}
\label{app:mel}

We briefly outline the modifications needed to the standard Melnikov
approaches
\cite{gh,ap,wiggins} relevant to Section~\ref{sec:mel}. Our system is
$ \dot{\mathbf z} = f \left( {\mathbf z} \right) + \epsilon \,
g \left( {\mathbf z} \right) $, as given in (\ref{eq:melset}).
Consider a particular parametrization of the heteroclinic $
\hat{\mathbf z}(\xi) $.  Imagine the perturbed system as embedded in
three-dimensional $ \left( {\mathbf z}, s \right) $ space.  In a
``time''-slice $ s = s_0 $, let $ {\mathcal T} $ be the normal vector
to the heteroclinic drawn at the point $ \hat{\mathbf z}(0) = {\mathbf
z}_0 $.  The usual approach is to compute the distance between the
perturbed manifolds measured along $ {\mathcal T} $, and this is
expandable as
\begin{equation}
\label{eq:distance1}
d(s_0,\epsilon) = \epsilon \, \frac{ M(s_0) }{ \left| {\mathbf f} \left( {\mathbf z}_0 \right)
\right| } + {\mathcal O}(\epsilon^2) \, .
\end{equation}
Let $ {\mathbf z}^u(s) $ be the trajectory of the perturbed flow which
intersects $ {\mathcal T} $ and which backwards asymptotes to the
perturbed fixed point $ {\mathbf a}(\epsilon) $.  In other words, $
{\mathbf z}^u(s) $ is a trajectory lying on $ {\mathbf a}(\epsilon)
$'s unstable manifold.  The standard approach \cite{gh,ap} is to
represent
\[
{\mathbf z}^\sigma (s) = \hat{\mathbf z} (s - s_0) + \epsilon \, {\mathbf z}_1^\sigma (s) +
{\mathcal O}(\epsilon^2)
\]
where $ \sigma = u $ (for ``unstable''), valid for $ - \infty < s \le
s_0 $ . A similar expansion on $ s_0 \le s < \infty $ with $ \sigma =
s $ (for ``stable''), works for the trajectory $ {\mathbf z}^s(s) $,
which intersects $ {\mathcal T} $ on the time-slice $ s_0 $ and which
lies on the stable manifold of the perturbed fixed point $ {\mathbf
b}(\epsilon) $.  Then, the standard Melnikov development (see
\cite{gh,ap}) allows the representation
\begin{equation}
\label{eq:distance2}
d(s_0,\epsilon) = \epsilon \, \frac{ \Delta^u(s_0) - \Delta^s(s_0) }{ \left| {\mathbf f} \left(
{\mathbf z}_0 \right) \right| } + {\mathcal O}(\epsilon^2) \, ,
\end{equation}
where
\[
\Delta^\sigma(s) = {\mathbf f} \left( \hat{\mathbf z}(s-s_0) \right) \wedge
{\mathbf z}_1^\sigma(s)
\]
for $ \sigma = u $ and $ \sigma = s $.  Now, \cite{gh,ap} derive
that
\begin{equation}
\label{eq:Delta}
\dot{\Delta}^\sigma = {\mathbf \nabla} \cdot {\mathbf f} \left( \hat{\mathbf z}(s-s_0) \right) \,
\Delta^\sigma + {\mathbf f} \left( \hat{\mathbf z}(s-s_0) \right) \wedge {\mathbf g} \left(
\hat{\mathbf z}(s-s_0),s \right) + {\mathcal O}(\epsilon)
\end{equation}
but since the unperturbed dynamical system is volume-preserving, have
the luxury of ignoring the first term on the right hand side.  We
cannot do so here, but we can neglect the second argument in $
{\mathbf g} $ since our case is autonomous.  To deal with the first
term, we multiply (\ref{eq:Delta}) by the integrating factor
\[
\mu(s) = \exp \left[ - \int_0^s {\mathbf \nabla} \cdot {\mathbf f} \left( \hat{\mathbf z}(r
- s_0) \right) \, dr \right]
\]
before proceeding.  Having done so, we integrate from $ - \infty $ to
$ s_0 $ by choosing $ \sigma = u $, then integrate from $ s_0 $ to $
\infty $ by choosing $ \sigma = s $, and then add the two equations to
get
\[
\Delta^u(s_0) - \Delta^s(s_0) = \int_{-\infty}^\infty \frac{\mu(s)}{\mu(s_0)} {\mathbf f}
\wedge {\mathbf g}  \left( \hat{\mathbf z}(s - s_0) \right) \, ds \, .
\]
(This is an adaptation of the standard process \cite{gh,ap}.)
In conjunction with (\ref{eq:distance1}) and (\ref{eq:distance2}), and also
employing the shift $ s - s_0 \rightarrow s $ in the integrand, we obtain the
Melnikov function
\[
M(s_0) = \int_{-\infty}^\infty \exp \left[ - \int_{0}^s {\mathbf \nabla} \cdot {\mathbf f} \left(
\hat{\mathbf z}(r) \right) \, dr \right] {\mathbf f} \wedge {\mathbf g}
\left( \hat{\mathbf z}(s) \right) ds \,
\]
which no longer depends on $ s_0 $.  Having dealt with the non
volume-preserving instance, the next step is to change our attitude:
rather than measuring the distance in a time-slice $ s $ but at a {\em
specific} point $ {\mathbf z}_0 $, we ignore time-slices (since our
perturbed system is itself autonomous), and allow the point to vary
along the heteroclinic.  To do so, choose a {\em different}
parametrization $ \hat{\mathbf w}(s) = \hat{\mathbf z}(s - \xi) $ of
the heteroclinic.  Thus, the point $ {\mathbf w}_0 = \hat{\mathbf
w}(0) =
\hat{\mathbf z}(-\xi) $ can be varied
along the heteroclinic by choosing different values of $ \xi $.
Therefore, $ \xi $ will represent different points along the
heteroclinic at which the distance measurement is to be made (cf.\
``heteroclinic coordinates'' of \cite{wiggins}).  Using the $ {\mathbf
w} $ trajectory, our earlier results can be expressed as
\begin{equation}
\label{eq:distance3}
d(s_0, \xi) = \epsilon \, \frac{ M(s_0,\xi )}{ \left| {\mathbf f}
\left( {\mathbf w}_0 \right) \right| } + {\mathcal O}(\epsilon^2)  =
\epsilon \, \frac{ M(\xi )}{ \left| {\mathbf f}
\left( {\mathbf z}(-\xi) \right) \right| } + {\mathcal O}(\epsilon^2), \, ,
\end{equation}
where
\begin{eqnarray*}
M(\xi) & = & \int_{-\infty}^\infty \exp \left[ - \int_0^s {\mathbf \nabla} \cdot
{\mathbf f} \left(
\hat{\mathbf w}(r) \right) dr \right] {\mathbf f} \wedge {\mathbf g} \left( \hat{\mathbf w}(s)
\right) ds \\
& = & \int_{-\infty}^\infty \exp \left[ - \int_0^s {\mathbf \nabla}
\cdot {\mathbf f} \left(
\hat{\mathbf z}(r - \xi) \right) dr \right] {\mathbf f} \wedge {\mathbf g}
\left( \hat{\mathbf z}(s - \xi)
\right) ds \\
& = & \int_{-\infty}^\infty \exp \left[ - \int_0^{s+\xi} {\mathbf \nabla} \cdot {\mathbf f} \left(
\hat{\mathbf z}(r - \xi) \right) dr \right] {\mathbf f} \wedge {\mathbf g} \left(
\hat{\mathbf z}(s)
\right) ds \\
& = & \int_{-\infty}^\infty \exp \left[ - \int_{-\xi}^{s} {\mathbf \nabla} \cdot {\mathbf f}
\left(
\hat{\mathbf z}(r) \right) dr \right] {\mathbf f} \wedge {\mathbf g} \left( \hat{\mathbf z}(s)
\right) ds \, .
\end{eqnarray*}
This, in conjunction with (\ref{eq:distance3}), is the expression used in Section~\ref{sec:mel}.

\section{Evans function definition}
\label{app:evans}

Here, we describe the Evans function approach for analyzing linear stability.
In general, the linear stability of a
localized traveling wave solution to a system of PDEs is obtained by
studying the eigenvalue problem
\begin{equation}
{\cal{L}} \, w=\lambda \, w \, ,
\label{eigenvalue}
\end{equation}
where the matrix differential operator ${\cal{L}}$ arises from the
linearization of the PDEs. The traveling solution is said to be
linearly stable if the spectrum of $\cal{L}$ lies in the closed left
half-plane.

The system (\ref{eigenvalue}) can be turned into a linear dynamical
system of the form
\begin{equation}
U_\xi={\bf A}(\xi,\lambda)\,U
\label{linear}
\end{equation}
where ${\bf A}(\xi,\lambda)$ is an $n\times n$ square matrix depending
on $\xi=x-c\,t$ and the spectral parameter $\lambda$ (in our case, $ n = 4$).
It can be shown that the asymptotic behavior of the solutions to (\ref{linear}) is
determined by the matrices
\[
{\bf A}_{\pm \infty}(\lambda)=\lim_{\xi\rightarrow \pm \infty}{\bf A}(\xi,\lambda)
\]
in the following sense (see \cite{BrDeGo02} for details). If $\mu^{+}$
(resp.\ $\mu^-$) is an eigenvalue of ${\bf A}_{+\infty}$ (resp.\ ${\bf
A}_{-\infty}$) with eigenvector $v^{+}$ (resp.\ $v^-$), then there
exists a solution $w^+$ (resp.\ $w^-$) to (\ref{linear}) with the
property that
\begin{equation}
\lim_{\xi\rightarrow \infty} w^{+}e^{-\mu^+\,\xi}=v^+\;\;\;
\left( {\mbox{resp.}}\;\;\lim_{\xi\rightarrow - \infty} w^{-}e^{-\mu^-\,\xi}=v^-
\right) \, .
\label{behaviorinf}
\end{equation}
Note that the superscript ``$+$'' refers to exponentially
decaying behavior at $\xi=+\infty$, while ``$-$''
refers to $\xi = -\infty$.

To study the linear stability, one should consider both the essential
and point spectrum of $\cal{L}$. The essential spectrum of ${\cal{L}}$
consists of the values of $\lambda$ for which ${\bf A}_\infty$ or
${\bf A}_{-\infty}$ has purely imaginary eigenvalues \cite{h}. The
point spectrum can be studied by means of the Evans function,
first introduced by Evans~\cite{Ev75} and later
generalized \cite{Al90}.  Roughly speaking, the zeros of this complex-valued function
are arranged to coincide with the point spectrum of $ \cal{L} $.

Let $\Omega$ denote a domain of the complex
$\lambda$ plane with no intersection with the essential spectrum and
let $n_s$ and $n_u$ denote, respectively, the number of eigenvalues of
${\bf A}_\infty$ with negative real part and the number of eigenvalues
of ${\bf A}_{-\infty}$ with positive real part in $\Omega$. We assume
that $n_s+n_u=n$. Let $w_i^+(\lambda,\xi)$, $i=1,2,...,n_s$ (resp.\
$w_i^-(\lambda,\xi)$, $i=1,2,...,n_u$) be linearly independent
solutions to (\ref{linear}) converging to zero as $\xi\rightarrow
\infty$ (resp.\ $\xi\rightarrow -\infty$) which are analytic of
$\lambda$ in $\Omega$. Clearly, a particular value of $\lambda$
belongs to the point spectrum of $\cal{L}$ if (\ref{linear}) admits a
solution that is converging to zero for both $\xi\rightarrow \pm
\infty$, that is if the space of solutions generated by the $w_i^+$
intersects with the one generated by the $w_i^-$. To detect such
values of $\lambda$ in $\Omega$, one can use the definition of the Evans
function given in \cite{sandstede}
\[
E(\lambda)=\det{\left(w_1^+,w_2^+,...w_{n_s}^+,w_1^-,w_2^-,...w_{n_u}^-\right)},
\]
in which the $w_i^\pm $ are evaluated at $\xi=0$.  This function is
analytic in $\Omega$, is real for real values of $\lambda$ and the
locations of the zeros of $E(\lambda)$ correspond to eigenvalues of
$\cal{L}$.

The first {\em numerical} computation of the Evans function was by
Evans himself in \cite{Ev77}, and followed by
\cite{SwEl90,EvansPego}. However, in all three papers $n_s=1$, in
which case a standard shooting argument can be used. In standard
shooting algorithms one follows the stable and/or unstable manifolds
at $ \xi=\pm \infty$. The Evans function is then given as the
intersection of these manifolds.  As shown in
Section~\ref{sec:stability}, our system has $ n = 4 $ and $ n_s = n_u
= 2 $.  This causes the following practical problem: although the
$n_s$ (or $n_u$ respectively) eigenvectors are linear independent
solutions of the eigenvalue problem (\ref{linear}) at $\xi=\pm
\infty$, the numerical integration scheme will lead to an inevitable
alignment with the eigen-direction corresponding to the largest
eigenvalue. This collapse of the eigen-directions is usually overcome
by using Gram-Schmidt orthogonalization. However, this is not
desirable for calculating the Evans function as it is a nonanalytic
procedure which then subsequently prohibits the use of Cauchy's
theorem (argument principle) to locate complex zeros of the Evans
function. The Evans function is therefore best calculated using
exterior algebra
\cite{NgR79,Br99,Br00,AfBr01,AlBr02,BrDeGo02,DeGo05,skms}.

We briefly review the method here, with specific regard to the situation in
which $ n = 4 $ and $ n_s = n_u = 2 $.  For more details the reader is
referred to \cite{AfBr01,AlBr02,BrDeGo02,DeGo05}, and to the numerical
computation in \cite{EvansWeberGroup}. The main idea behind
exterior algebra methods (or compound matrices methods) is that the
linear system (\ref{linear}) induces a dynamical system on the
wedge-space~$\CwedgeX$ for $n_s=n_u=2$. The wedge-space~$\CwedgeX$ is
the space of all two forms on ${\mathbb C}^{n}$. This is a space of
dimension ${4\choose 2}=6$. The induced
dynamics on the wedge-space~$\CwedgeX$ can be written as
\begin{equation}\label{Unk}
{\bf U}_\xi = {\bf A}^{(2)} (\xi){\bf U} , \quad {\bf U}\in\CwedgeX.
\end{equation}
Here the linear operator (matrix) ${\bf A}^{(2)}$ is the restriction
of ${\bf A}(\xi,\lambda)=\{a_{ij}\}$ to the wedge space $\CwedgeX$. Using the standard
basis of $\CwedgeX$
\begin{equation}\label{5.1}
\begin{array}{l}
{\bf \omega}_1 = {\bf e}_1\wedge{\bf e}_2 ,\quad
{\bf \omega}_2 = {\bf e}_1\wedge{\bf e}_3 ,\quad
{\bf \omega}_3 = {\bf e}_1\wedge{\bf e}_4 ,\\
{\bf \omega}_4 = {\bf e}_2\wedge{\bf e}_3 ,\quad
{\bf \omega}_5 = {\bf e}_2\wedge{\bf e}_4 ,\quad
{\bf \omega}_6 = {\bf e}_3\wedge{\bf e}_4 ,
\end{array}
\end{equation}
where ${\bf e}_{1,2,3,4}$ is the standard basis of ${\mathbb C}^{n}$,
we can find the matrix ${\bf A}^{(2)}:\CwedgeX\to\CwedgeX$ as a
complex $6\times 6$ matrix. With respect to the basis (\ref{5.1}),
${\bf A}^{(2)}$ takes the explicit form
\[
{\bf A}^{(2)}=
{\small
\left[
\begin{array}{ccccccccccc}
a_{11}\!+\! a_{22} && a_{23} && a_{24} && -a_{13} && -a_{14} && 0\\
a_{32} && a_{11}\!+\! a_{33} && a_{34} && a_{12} && 0 && -a_{14}\\
a_{42} && a_{43} && a_{11}\!+\! a_{44} && 0 && a_{12} && a_{13}\\
-a_{31}&& a_{21}&& 0 && a_{22} \!+\! a_{33} && a_{34} && -a_{24}\\
-a_{41}&& 0 && a_{21} && a_{43} && a_{22} \!+\! a_{44} && a_{23}\\
0  && -a_{41}&& a_{31}&& -a_{42} && a_{32} && a_{33} \!+\! a_{44}
\end{array}
\right]
}
\]
General aspects of the numerical implementation of this theory and
details for these constructions in more general systems can be found
in \cite{AlBr02,BrDeGo02}.

Linearity assures that the induced matrix ${\bf A}^{(2)}(\xi,\lambda)$
is also differentiable and analytic. Hence, the limiting matrices,
\[
{\bf A}_{\pm \infty}^{(2)}(\lambda) =
\lim_{\xi\to\pm\infty}{\bf A}^{(2)}(\xi,\lambda)\,,
\]
will exist. It can readily be shown that the eigenvalues of the matrix
${\bf A}_{\pm \infty}^{(2)}(\lambda)$ consists of all possible sums of
2~eigenvalues of ${\bf A}_{\pm\infty}(\lambda)$. Therefore, for
$\Re(\lambda)>0$, the eigenvalue of ${\bf A}_{+\infty}^{(2)}(\lambda)$
with the most negative real part is given by $\sigma_+(\lambda) =
\mu_1^+ + \mu_2^+$. The eigenvalue $\sigma_+(\lambda)$ has real part
strictly less than any other eigenvalue of ${\bf
A}_{+\infty}^{(2)}(\lambda)$. Analogously, the eigenvalue of ${\bf
A}_{-\infty}^{(2)}(\lambda)$ with the largest non-negative real part
is given by $\sigma_-(\lambda) = \mu_1^- + \mu_2^-$. The eigenvalue
$\sigma_-(\lambda)$ has real part strictly greater than any other
eigenvalue of ${\bf A}_{-\infty}^{(2)}(\lambda)$. Note that the
eigenvalues $\sigma_\pm$ are simple, and analytic functions of
$\lambda$.

Let $\zeta^\pm(\lambda)$ be the eigenvectors associated with
$\sigma_\pm(\lambda)$, defined by
\begin{equation}\label{eigVs}
{\bf A}_{+\infty}^{(2)}(\lambda)\zeta^+(\lambda) =
\sigma_+(\lambda)\zeta^+(\lambda)
\quad {\rm {and}} \quad
{\bf A}_{-\infty}^{(2)}(\lambda)\zeta^-(\lambda) =
\sigma_-(\lambda)\zeta^-(\lambda)\,.
\end{equation}
These vectors can always be constructed in an analytic way (see
\cite{BrDeGo02}), and are readily found to be $\zeta^\pm(\lambda) =
v_1^\pm \wedge v_2^\pm$.

Let ${\bf U}^{\pm}(\xi,\lambda) \in \CwedgeX$ be the solution of the
linear system (\ref{Unk}) which satisfy $\lim_{\xi\to\pm\infty}
e^{-\sigma_\pm(\lambda)\xi} {\bf U}^\pm(\xi,\lambda) =
\zeta^\pm(\lambda)$. This allows us to define the Evans function as
\begin{equation}\label{2.9}
E(\lambda) = {\cal{N}} \, {\bf
U}^-(\xi,\lambda)\wedge {\bf U}^+(\xi,\lambda)\,,\quad
\lambda\in\Lambda\,,
\end{equation}
where
\begin{equation}\label{2.10}
{\cal{N}} = \re^{-\int_0^\xi\tau(s,\lambda)ds} \quad {\rm{and}} \quad
\tau(\xi,\lambda) = {\rm Tr}( {\bf A}(\xi,\lambda) ) \; .
\end{equation}
Expressing ${\bf U}^\pm(\xi,\lambda)$ as a linear combination with
respect to the basis (\ref{5.1})
\[
{\bf U}^\pm(\xi,\lambda)  = \sum_j^6U_j^{\pm} \omega_j\; ,
\]
the expression for the Evans function (\ref{2.10}) can be simplified
to
\begin{equation}\label{UVmatch1}
E(\lambda) = {\cal{N}} \, \lbk{{\bf U}^-(\xi,\lambda)},
{\bf \Sigma} {\bf U}^+(\xi,\lambda)\rbk_6\,,
\end{equation}
where $\lbk\cdot , \cdot\rbk_6$ is the complex inner product in
$\C^4$, and the representation of the Hodge-star operator ${\bf
\Sigma}$ in the basis (\ref{5.1}) is
\[
{\bf \Sigma} =
\left[
\begin{array}{ccccccccc}
0 & \hfill0 & \hfill0 & \hfill0 & \hfill0 & \hphantom{-}1 \\
0 & \hfill0 & \hfill0 & \hfill0 & -1      & 0 \\
0 & \hfill0 & \hfill0 & \hphantom{-}1 & \hfill0 & 0 \\
0 & \hfill0 & \hphantom{-}1  & \hfill0 & \hfill0 & 0 \\
0 & -1 & 0 & \hfill0 & \hfill0 & 0 \\
\hphantom{-}1 & \hfill0 & \hfill0 & 0 & \hfill0 & 0
\end{array}
\right] \; .
\]
Using the Hodge-star operator, we can relate the the most unstable
solution~${\bf U}^-$ of the linearized system at $\xi=-\infty$ with
the most unstable solution of the adjoint system of (\ref{Unk}) at
$\xi=-\infty$. Details can be found
in~\cite{BrDe99,AlBr02,BrDeGo02}. This suggests a normalization of the
asymptotic eigenvectors according to
\begin{equation}\label{normalisation}
\lbk{{\zeta}^-},
{\bf \Sigma} {\zeta}^+\rbk_6 = 1\; ,
\end{equation}
which assures that $E(\lambda)\to 1$ for $|\lambda| \to \infty$.

Note that the translational invariance of (\ref{eq:governpde})
guarantees that the Evans function can be evaluated at any (fixed)
spatial location $\xi^\star$. However, to avoid unwanted growing of the
solutions ${\bf U}^{\pm}$ we will consider the scaled solutions
\begin{equation}\label{scalep}
\widetilde{\bf U}^\pm(\xi,\lambda) = e^{-\sigma_\pm(\lambda)\xi}
{\bf U}^\pm(\xi,\lambda) \; .
\end{equation}
The scaling (\ref{scalep}) ensures that $\widetilde{\bf
U}^+(\xi,\lambda)\big|_{\xi=\xi^\star}$ is bounded. The corresponding
equation on $\CwedgeX$
\begin{equation}\label{xplus}
\frac{d\ }{d \xi}\widetilde{\bf U}^\pm
= [{\bf A}^{(2)}(\xi,\lambda) - \sigma_\pm(\lambda){\bf I}]
\widetilde{\bf U}^\pm\,,
\quad \widetilde{\bf U}^\pm(\xi,\lambda)\big|_{\xi=L_{\pm \infty}} =
\zeta^\pm(\lambda)\,,
\end{equation}
is integrated from $\xi=L_{\pm \infty}$ to $\xi=\xi^\star$ (where
$\xi^\star$ is arbitrary but fixed).\\

The system (\ref{xplus}) can be integrated using the second-order implicit
midpoint method.  For a system in the form ${\bf U}_\xi = {\bf
B}(\xi,\lambda) {\bf U}$, each step of the implicit midpoint rule takes
the form
\begin{equation}\label{glrk2}
{\bf U}^{n+1} = [
{\bf I} - \fr\Delta x{\bf B}_{n+1/2}]^{-1}
[{\bf I} + \fr\Delta x{\bf B}_{n+1/2}]\,{\bf U}^{\,n}\; ,
\end{equation}
where ${\bf B}_{n+1/2} = {\bf B}(x_{n+1/2},\lambda)$.

\bibliographystyle{siam}
\bibliography{newcombust1}

\end{document}